\documentclass[10pt, letterpaper, onecolumn, oneside]{article}

\usepackage[noend]{algorithmic}
\usepackage{algorithm}
\usepackage{amssymb,amsmath}
\usepackage[dvips]{graphicx,epsfig}
\usepackage{subfigure}
\usepackage{balance}
\newtheorem{theo}{\bf Theorem}
\newtheorem{defn}{\bf Definition}

\input{epsf}
\begin{document}

\title{PRESS: A Novel Framework of Trajectory Compression in Road Networks}

\author{
Renchu Song$^\dagger$$^\star$ \ \ Weiwei Sun$^\dagger$$^\star$ \ \ Baihua Zheng$^\ddagger$\ \ Yu Zheng$^\S$ 
\\
\begin{tabular}{ccc}
\normalsize $^\dagger$Fudan University, Shanghai, China, \{songrenchu, wwsun\}@fudan.edu.cn \\
\normalsize $^\star$Shanghai Key Laboratory of Data Science, Fudan University, Shanghai, China \\
\normalsize $^\ddagger$Singapore Management University, Singapore, bhzheng@smu.edu.sg \\
\normalsize $^\S$Microsoft Research, Beijing, China, yuzheng@microsoft.com \\
\end{tabular}
}

\maketitle


\begin{abstract}
Location data becomes more and more important. In this paper, we focus on the trajectory data, and propose a new framework, namely PRESS (\emph{\underline{P}aralleled \underline{R}oad-Network-Based Trajectory Compr\underline{ess}ion}), to effectively compress trajectory data under road network constraints. Different from existing work, PRESS proposes a novel representation for trajectories to separate the spatial representation of a trajectory from the temporal representation, and proposes a \emph{Hybrid Spatial Compression} (HSC) algorithm and error \emph{Bounded Temporal Compression} (BTC) algorithm to compress the spatial and temporal information of trajectories respectively. PRESS also supports common spatial-temporal queries without fully decompressing the data. Through an extensive experimental study on real trajectory dataset, PRESS significantly outperforms existing approaches in terms of saving storage cost of trajectory data with bounded errors.
\end{abstract}

\section{Introduction}

The advance in location-acquisition technologies has led to a huge volume of spatial trajectories, e.g., the GPS trajectories of vehicles, each of which is comprised of a sequence of time-ordered spatial points. As the trajectories are in huge volume and some points in a trajectory are redundant, application systems on trajectories have to bear high communication loads and expensive data storage. This is calling for trajectory compression technologies that can reduce the storage cost while keeping the utility of a trajectory. In this paper, we propose a trajectory compression framework under the road network constraints, namely PRESS (\emph{\underline{P}aralleled \underline{R}oad-Network-Based Trajectory Compr\underline{ess}ion}). The main objective is to achieve a spatial lossless and temporal error-bounded compression, and meanwhile provide support to popular LBS applications.

\begin{figure}[htb]%
        \centering
        \psfig{file=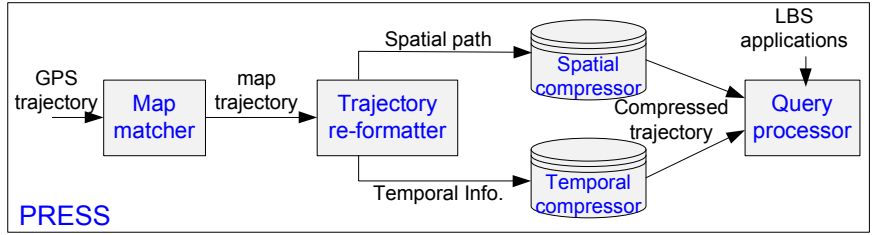,width=0.7\textwidth}
        \vspace{-0.15in}
        \caption{PRESS framework}
        \label{fig:press}
\end{figure}

The PRESS framework has five components, namely \emph{map matcher}, \emph{trajectory re-formatter}, \emph{spatial compressor}, \emph{temporal compressor}, and \emph{query processor}, as shown in Fig.~\ref{fig:press}. Taking raw GPS trajectories as input, map matcher maps each trajectory into a sequence of edges in the road network, which will be reformatted into a spatial path and a temporal sequence via trajectory re-formatter. Thereafter, the compression takes place in parallel. The spatial path is compressed by spatial compressor based on \emph{Hybrid Spatial Compression} (HSC) algorithm; and the temporal sequence is compressed by temporal compressor based on \emph{Bounded Temporal Compression} (BTC) algorithm. The compressed spatial path and compressed temporal sequence are then passed to query processor to support different application needs.

\vspace{-0.1in}


Different from existing works, we consider both the compression ratio and the utility of the compressed trajectories. In general, the higher the compression ratio, the lower the quality of the compressed trajectory, which directly affects data utility. Consequently, it is challenging to propose a novel approach to achieve a high compression ratio with high quality compressed trajectories, especially under road network constraints.

PRESS tackles this issue from three different angles. First, it observes that the spatial path and the temporal information of a trajectory have different features and hence it strategically separates the spatial path from the temporal information when presenting a trajectory. The clear separation allows us to compress the spatial path and the temporal information separately. Second, a lossless spatial compression algorithm HSC is proposed to effectively compress the spatial path using significantly less space without losing any spatial information. It has two stages.
%
%
The first stage compression is based on shortest paths. Given a sub-trajectory $T_{sub}$ from edges $e_i$ to $e_j$, if $T_{sub}$ is exactly the same as the shortest path from $e_i$ to $e_j$, $T_{sub}$ will be replaced by ($e_i$, $e_j$). As in many cases we tend to take shortest paths in real life, this compression can effectively reduce the number of edges we have to maintain for each trajectory. The second stage compression is based on frequent sub-trajectory (FST) coding. The main idea is to decompose a trajectory into a sequence of FSTs, each of which is represented by a unique code (e.g., Huffman code). The more popular the FST, the shorter the corresponding code and the more the space savings.  Meanwhile, PRESS designs a temporal compression algorithm BTC to compress temporal information with bounded errors. BTC is very flexible, and it can compress the temporal information based on the error bounds specified by different applications. As a summary, the lossless nature of the spatial compression and the error-bounded nature of the temporal compression guarantee the high quality of the compressed trajectories. Last but not the least, PRESS also supports many popular spatial-temporal queries commonly used in location-based services (LBSs) such as $where_{at}$, $when_{at}$ and $range$ queries without fully recovering the compressed trajectories.

An extensive experimental study has been conducted on a real trajectory dataset to validate the effectiveness and efficiency of PRESS. According to the results, PRESS can save up to $78.4\%$ of the original storage cost. Let $|T|$ be the length of a trajectory $T$. Both HSC and BTC have the compression time complexity of $O(|T|)$, and hence the compression time complexity of PRESS is $O(|T|)$. As compressed temporal sequences share the same format as original ones, BTC does not require any decompression process. In other words, the decompression time complexity of PRESS is equal to that of HSC, i.e., $O(|T|)$. In addition, PRESS can significantly accelerate spatial-temporal queries. In brief, PRESS outperforms the state-of-the-art approaches in terms of the compression ratio, the time consumption and the acceleration of spatial-temporal queries.

The rest of the paper is organized as follows.
Section~\ref{sec:tra-rep} presents our new approach to represent a trajectory. Section~\ref{sec:spatial-compression} and Section~\ref{sec:temporal-compression} introduce the detailed spatial compression and temporal compression respectively. Section~\ref{sec:application} explains how to support some common queries via compressed trajectories. Section~\ref{sec:exp} presents our experimental studies. Section~\ref{sec:related} reviews related work. Finally, Section~\ref{sec:conclusion} concludes this paper with some directions for future work.





\section{Trajectory Representation}
\label{sec:tra-rep}

In our work, a road network is defined as a directed graph $G = (V, E)$, where $V$ is the vertex set and $E$ is the edge set. The weight on an edge $e$, denoted as $w(e)$, can be physical distance, travel time or other costs according to different application context. A trajectory is the path that a moving object follows through space as a function of time. Consequently, it contains both spatial information and temporal information. Traditional approaches represent trajectories via a sequence of $n$ triples in the form of $((x_1, y_1, t_1)$, $(x_2, y_2, t_2)$, $\cdots$, $(x_n, y_n, t_n))$, where $(x_i, y_i)$ is the position in the 2D Euclidean space at time stamp $t_i$.

We propose a different representation of trajectories in the road network. Instead of combining positions and time stamps together like existing approaches do, we separate the locations from time stamps. In other words, a trajectory is represented by a spatial path and a temporal sequence. This clear separation enables us to design different compression approaches for spatial information and temporal information respectively, so that both spatial compression and temporal compression can achieve high compression effectiveness without constraining each other. In the following, we will explain how to represent the spatial information and temporal information via spatial path and temporal sequence, respectively.

\begin{figure}[htb]%
        \centering
        \psfig{file=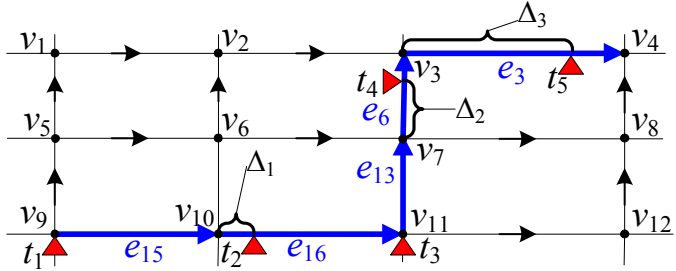,width=0.55\textwidth}
        \vspace{-0.2in}
        \caption{Sample trajectory in a road network}
        \label{fig:sample-tra}
\end{figure}

The spatial path of a trajectory in a road network is a sequence of consecutive edges. As shown in Fig.~\ref{fig:sample-tra}, a trajectory sequentially passes edges $e_{15}$, $e_{16}$, $e_{13}$, $e_{6}$, and $e_{3}$. Consequently, it can be represented by a spatial path, in the format of $\langle e_{15}, e_{16}, e_{13}, e_{6}, e_{3}\rangle$. Note trajectories can start from and/or end at any point of an edge, not necessarily an endpoint. For example, the example path ends at a point along edge $e_3$. We will tackle this issue via the temporal sequence presented in the following.

\vspace{-0.12in}
\vspace{-0.1in}


The temporal information of a trajectory defines the time when an object locates at a specific location. For example, the triple $(x_i, y_i, t_i)$ used in the traditional representation tells that the object is located at position $(x_i,y_i)$ at time stamp $t_i$. However, this representation does not facilitate the spatial queries in road networks. Consider a common query that asks for the average moving speed of an object $obj$ during a period [$t_i, t_j$] with $1\le i<j\le n$. Positions $(x_i, y_i)$ and $(x_j, y_j)$ do not capture any distance information and we have to explore the road network to calculate the distance traveled by $obj$ from $t_i$ to $t_j$. Consequently, we propose to use the tuple $(d_i, t_i)$ to capture the temporal information. To simplify the discussion, $d_i$ in this paper represents the network distance the object has traveled at the time stamp $t_i$ since the start of the trajectory. More generally, $d_i$ can represent other weight information of the edges, e.g., travel time or other costs based on application needs.

\begin{figure}[htb]%
\vspace{-0.12in}
        \centering
        \psfig{file=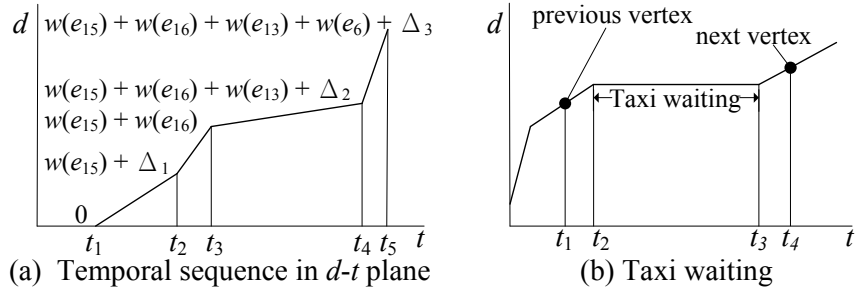,width=0.7\textwidth}
        \vspace{-0.17in}
        \caption{Temporal sequence}
        \label{fig:taxi-waiting}
\vspace{-0.13in}
\end{figure}

Back to the example trajectory shown in Fig.~\ref{fig:sample-tra}. There are five time stamps denoted as $t_1$, $t_2$, $t_3$, $t_4$, and $t_5$, respectively. Based on our newly proposed temporal sequence representation, the temporal information of our example trajectory will be represented by five tuples. They are $\langle 0, t_1\rangle$, $\langle w(e_{15})+\Delta_1, t_2\rangle$, $\langle w(e_{15})+w(e_{16}), t_3\rangle$, $\langle w(e_{15})+w(e_{16})+w(e_{13})+\Delta_2, t_4\rangle$, and $\langle w(e_{15})+w(e_{16})+w(e_{13})+w(e_{6})+\Delta_3, t_5\rangle$, as shown in Fig.~\ref{fig:taxi-waiting}(a). The first tuple means the object starts the trajectory at time stamp $t_1$ and the corresponding distance it has traveled since start is zero (i.e., $d_1=0$), the second tuple means the object has traveled $w(e_{15})+\Delta_1$ distance at time stamp $t_2$ with $w(e_{15})$ representing the distance of edge $e_{15}$, and so on. Note the last tuple does not locate at any endpoint, and the same can happen to the first tuple too.


Although we are not the first one to represent the spatial path of a trajectory via edges, we want to highlight that our approach to separating the spatial information from the temporal information when representing trajectories is very \emph{unique}. Former approach~\cite{JSS13} uses the vertices in a road network (i.e., the endpoints of edges) to capture the spatial information of trajectories, together with the time when the object passes those vertices. However, the time stamps when the object passes those vertices do not cover the entire temporal information. For example, a taxi might stop for a long time somewhere between two vertices. By retaining two time stamps of the vertices, we can only assume that the taxi drives at a low uniform speed on the edge, which is not the real case. Our approach can easily tackle this issue. As illustrated in Fig.~\ref{fig:taxi-waiting}(b), we understand that the taxi moves slowly from $t_1$ to $t_2$, gets stuck from $t_2$ to $t_3$, and then moves slowly again from $t_3$ to $t_4$.

\section{Spatial Compression}
\label{sec:spatial-compression}

After presenting the formal representation of trajectories, we are ready to present \emph{Hybrid Spatial Compression} (HSC), the two-stage spatial compression algorithm. It takes an initial spatial trajectory $T=\langle e_1, e_2, \cdots, e_n\rangle$ as an input, and performs shortest path compression on the first stage and then frequent sub-trajectory compression on the second stage. Existing works use $n$ original sampled positions to keep track of a trajectory, and propose to use $m$ ($< n$) positions for capturing the trajectory in order to cut down the storage cost. However, all the existing approaches based on this idea cannot \emph{fully} capture the spatial path traveled by the trajectory. Consider our sample trajectory. Its spatial path can be represented by six vertices, i.e., $v_9$, $v_{10}$, $v_{11}$, $v_7$, $v_3$, and $v_4$. If we reduce the vertex number from original six to three and represent the trajectory by $v_9$, $v_{11}$, and $v_4$, we can only tell that in this trajectory, object $obj$ moves from $v_9$ to $v_4$ via vertex $v_{11}$ but we cannot tell how the object $obj$ moves from $v_9$ to $v_{11}$, and then from $v_{11}$ to $v_4$. Consider the movement from $v_{11}$ to $v_4$, the object $obj$ could take path ($v_{11}$, $v_{12}$, $v_8$, $v_4$), or ($v_{11}$, $v_{7}$, $v_{3}$, $v_4$), or ($v_{11}$, $v_{7}$, $v_8$, $v_4$), which is uncertain.

Although existing works propose various metrics to guarantee the similarity between compressed trajectories and original ones, \emph{none} is error-free. They trade in the accuracy of trajectories' spatial information for the saving of storage cost. Alternatively, our two-stage HSC approach is \emph{error-free}. The compressed trajectory returned by HSC, although taking less space, captures the spatial path of the original trajectory as it is. In HSC, we make two assumptions. i) Objects tend to take the shortest path instead of longer ones in most if not all cases; and ii) the trajectories are not uniformly distributed in the road network and there are certain edge sequences which are passed through more frequently.

\subsection{Shortest path compression}

Given a source $s$ and a destination $e$, most of the time we will take the shortest path (SP) between $s$ and $e$ if all the edges roughly share the similar traffic condition. Under this assumption, we can predict that most, if not all, of the trajectories consist of a sequence of shortest paths. Our first compression is motivated by this observation, and takes full advantage of shortest paths.

\begin{figure}[htb]%
\vspace{-0.1in}
        \centering
        \psfig{file=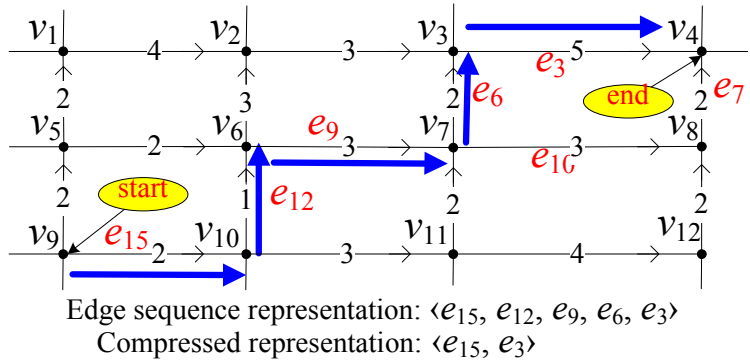,width=0.6\textwidth}
        \vspace{-0.15in}
        \caption{ Example of shortest path compression}
        \label{fig:sp-compress}
\vspace{-0.1in}
\end{figure}

We assume that all-pair shortest path information is available via a pre-processing of the road network. This can be achieved by any of the well known shortest path algorithms. If there are several shortest paths between a pair of edges, we only record one of them to eliminate any ambiguity during compression.
%
We assume $SP(e_i, e_j)$ denotes the shortest path from edge $e_i$ to edge $e_j$, and maintain a structure $SP_{end}(e_i, e_j)$ recording the last edge (the edge right before $e_j$) of $SP(e_i, e_j)$ for each pair of edges. Take the partial road network shown in Fig.~\ref{fig:sp-compress} as an example. Assume the number in the middle of each edge indicates the network distance of the edge, then $SP(e_{15}, e_7) = \langle e_{15}, e_{12}, e_9, e_{10}, e_7\rangle$, $SP_{end}(e_{15}, e_7)$ $= e_{10}$, $SP_{end}(e_{15}, e_{10}) = e_{9}$, and so on.

Before we present the detailed algorithm, we use a running example to illustrate the SP compression. The main idea is to skip the detailed sub-trajectory $\langle e_i, e_{i+1}, \cdots, e_j\rangle$ if it matches exactly the shortest path from $e_i$ to $e_j$, i.e., replacing $SP(e_i, e_j)$ with $e_i$ and $e_j$ only. As shown in Fig.~\ref{fig:sp-compress}, the original trajectory $T=\langle e_{15}, e_{12}, e_9, e_6, e_3\rangle$. Initially, the SP compression algorithm enrolls the first edge $e_{15}$ into $T'$. Thereafter, it scans the subsequent edges one by one. For the second edge $e_{12}$, $e_{15}$ and $e_{12}$ are adjacent and the process continues.
For the third edge $e_9$, $SP_{end}(e_{15}, e_9) = e_{12}$, so edge $e_{12}$ can be skipped. For the fourth edge $e_6$, $SP_{end}(e_{15}, e_6) = e_9$ and hence $e_9$ is also skipped. Next, $SP_{end}(e_{15}, e_3) = e_6$, so $e_{6}$ is skipped. Finally, the algorithm enrolls the last edge $e_3$ into $T'$ to finish the process and replaces $T$ with $T'=\langle e_{15}, e_3\rangle$. As it scans each edge in $T$ once, its complexity is $O(|T|)$.


The pseudo code is listed in Algorithm~\ref{algo:shortestpath}. It takes a trajectory $T=\langle e_1, e_2,\cdots, e_n\rangle$ as an input. It enrolls the first edge $e_1$ into the compressed trajectory $T'$, uses a pointer $index$ to record the tail edge of $T'$, and then sequentially scans the remaining edges $e_i$. If the trajectory $\langle e_{index}, e_{index+1}, \cdots, e_i\rangle$ matches exactly the shortest path from $e_{index}$ to $e_i$, there is no need to record $e_i$ into $T'$. After scanning $e_2$ until $e_{n-1}$, the algorithm terminates by returning $T' \cup e_n$. The main idea of SP compression is to replace the shortest path between two pair of edges with those two edges, and there are multiple ways to implement it. The SP algorithm proposed in this work is based on greedy algorithm, and it actually generates the largest compression ratio during shortest path compression, as stated in Theorem~\ref{theo:SP-containment}.
%
%
\begin{algorithm}[htb]
\begin{small}
\caption{Shortest Path Compression}\label{algo:shortestpath}
{\bf{Input}}: a road network $G=(V,E)$, a trajectory $T=\langle e_1, e_2,\cdots, e_n\rangle$;\\
{\bf{Output}}: a compressed trajectory $T'$;\\
\hspace*{0.03in} {\bf Procedure:}
\begin{algorithmic}[1]
\STATE $T' \gets e_1$; $index \gets 1$;\\
\FOR{$i\gets 2$ to ${n-1}$}
    \IF{$SP_{end}(e_{index}, e_{i+1}) \ne e_i$}
        \STATE $Append(T', e_i)$, $index \gets i$;\\
    \ENDIF
\ENDFOR
\STATE \textbf{return} $Append(T', e_n)$;\\
\end{algorithmic}
\end{small}
\end{algorithm}

\begin{theo}\label{theo:SP-containment}
The greedy algorithm is the optimal algorithm resulting in the largest compression ratio during SP compression.
\end{theo}

\noindent
{\bf{Proof.} }

Assume the input trajectory $T = \langle e_i, \cdots, e_j\rangle$. Now we prove that our greedy SP compression actually generates the optimal solution in terms of the number of edges by induction on $n$, the length of the compressed trajectory.

When $n = 3$, $T' = \langle e_i, e_m, e_j\rangle$. If the output $T'$ is not optimal, there must be a compressed trajectory $T''$ with length $< |T'| (=3)$. As the starting edge and the ending edge of $T$ must be preserved, $T'' = \langle e_i, e_j\rangle$. In other words, $T$ passes exactly the shortest path from $e_i$ to $e_j$. Based on SP compression, it will not output $T' = \langle e_i, e_m, e_j\rangle$ if $T = SP(e_i, e_j)$. Our statement is true when $n=3$.

Now, we assume the statement is true for all $k < n$ and let $T’$ be a compressed trajectory of length $n$ returned by our SP compression in the form of $\langle e_i, e_m, \dots , e_{m+n-3}, e_j\rangle$. If our statement is not true, there must be another compressed trajectory $T''$ with length $|T''| < |T' | (= n)$. If $T''$ and $T'$ share at least one common edge in addition to $e_i$ and $e_j$ (i.e., $\exists  e_k \in T'' \cap T'$ with $k \in [m, m+n-3]$), we can decompose $T'$ into $T'_1 = \langle e_i, e_m, \dots, e_k\rangle$  and $T'_2 = \langle e_k, \dots, e_j\rangle$. As $| T'_1|< n$ and $| T'_2|< n$, they both must be optimal solutions and our assumption that $|T''|<|T'|$ is not true. Otherwise, $T''$ and $T'$ do not share any common edge except $e_i$ and $e_j$. Without loss of generality, we can decompose $T'$ into $T'_1 = \langle e_i, e_m, \dots, e_k\rangle$  and $T'_2 = \langle e_k, e_{k+1}, \dots, e_j\rangle$ such that there is an edge $e'_k \in T''$ and $e'_k$ locates after $e_k$ but before $e_{k+1}$ in the original trajectory $T$. Accordingly, $T''$ can be decomposed into two sub-trajectories by $e'_k$ with $T''_1 = \langle e_i, \dots , e'_k\rangle$ and $T''_2 = \langle e'_k, \dots , e_j\rangle$. On the other hand, the original trajectory $T$ can be decomposed into two sub-trajectories $T_1$ and $T_2$ at edge $e'_k$ with $T_1 = \langle e_i, \dots , e'_k\rangle$, and $T_2 = \langle e'_k, \dots , e_j\rangle$. For $T_1$, the compressed trajectory $T'_1$ returned by the greedy algorithm must be $T'_1$ followed by $e'_k$, and $T''_1$ represents another possible compressed form returned by other SP compression. As $| T'_1| + 1\le n$, it is guaranteed that $|T'_1 |+1\le |T''_1|$. For $T_2$, we are certain the compressed form returned by PRESS will be $T'_2$ but started with $e'_k$ not $e_k$, as guaranteed by the SP-containment property; and as $| T'_2|\le n$, it is guaranteed that $| T'_2 |\le |T''_2|$. That is to say, $| T'_1 |+1 + | T'_2 | \le |T''_1| + |T''_2|$. As $|T'| = |T'_1| + |T'_2|-1$ and $|T''| = |T''_1| + |T''_2|-1$, we have $|T'| = | T'_1 |+| T'_2|-1< | T'_1 |+| T'_2| \le |T''_1| + |T''_2|-1 =|T''|$ which contradicts our assumption that $|T''|<|T'|$. Our assumption is invalid and the proof completes. $\blacksquare$
%

The decompression process is straightforward. Given a compressed trajectory $T' =\langle e_1, e_2, \cdots, e_m\rangle$, we sequentially scan each pair of edges ($e_i$, $e_{i+1}$). If they are not adjacent, we complement the trajectory with the shortest path $SP(e_i, e_{i+1})$. As $(e_{15}, e_3)$ is the only pair and $e_{15}$ is not adjacent to $e_3$, we complement $T$ with $SP(e_{15}, e_3)=\langle e_{15}, e_{12}, e_9$, $e_6, e_3\rangle$. In order to obtain $SP(e_i, e_j)$, we only need to visit $SP_{end}(e_i, e_j)$, $SP_{end}(e_i, SP_{end}(e_i, e_j))$ and so on. This step takes as many times as the length of the shortest path and hence the time complexity of the decompression process is also $O(|T|)$.

\subsection{Frequent sub-trajectory compression}

As we assume previously, trajectories are not evenly distributed within the road network and edges in a road network are not accessed uniformly. In other words, certain edge sequences are much more popular than others in terms of frequency. If we are able to locate the very popular sub-trajectories, named \emph{frequent sub-trajectory} (FST), then we can use certain coding scheme to compress them and to replace them in the trajectories with the corresponding codes. 

Given a large set of trajectory data, the concept of FST makes sense. Consequently, the compression based on FST is not effective if the underlying dataset is small. In addition, we also assume the trajectory dataset is periodical. That means if we collect all the trajectories of all the cars moving within a city for a duration of several months, the dataset of one day should be similar to the dataset of another day. Under this assumption, we can locate FSTs based on a subset of the complete trajectory dataset, which corresponds to the training process in data mining. Note that the input training dataset is a subset of the complete trajectory dataset after the SP compression. For example in our experiments, we take the trajectories corresponding to one day as a training dataset, perform SP compression for each trajectory in the training dataset, and then pass them to the second stage for FST mining. In the following, we explain how to mine FSTs, how to decompose a trajectory based on the mined FSTs, and the detailed decoding process, the three main steps of FST compression.

\subsubsection{FSTs mining}

As we want to compress the trajectories based on FSTs, we have to locate all the FSTs first. The problem of mining FSTs is similar to the frequent pattern mining problem~\cite{han2006data,luo2013finding,wong2006mining} in data mining. 
In this work, we propose a novel approach to locate FSTs. 
We treat sub-trajectories as \emph{strings} and use Huffman coding~\cite{IRE52_1098} to compress them. The more frequent a sub-trajectory is (based on its frequency in the training set), the shorter the corresponding code is and hence more savings in terms of storage cost are expected when compressing the trajectories. The basic idea is to first build a \emph{Trie}~\cite{kauth1973art} based on the training set to represent sub-trajectories as strings, next form an Aho-Corasick automaton~\cite{ACMC75_333} to enable a decomposition of a trajectory into a set of sub-trajectories, and then use Huffman coding to compress the trajectories.

\begin{figure}[htb]%
\vspace{-0.1in}
        \centering
        \psfig{file=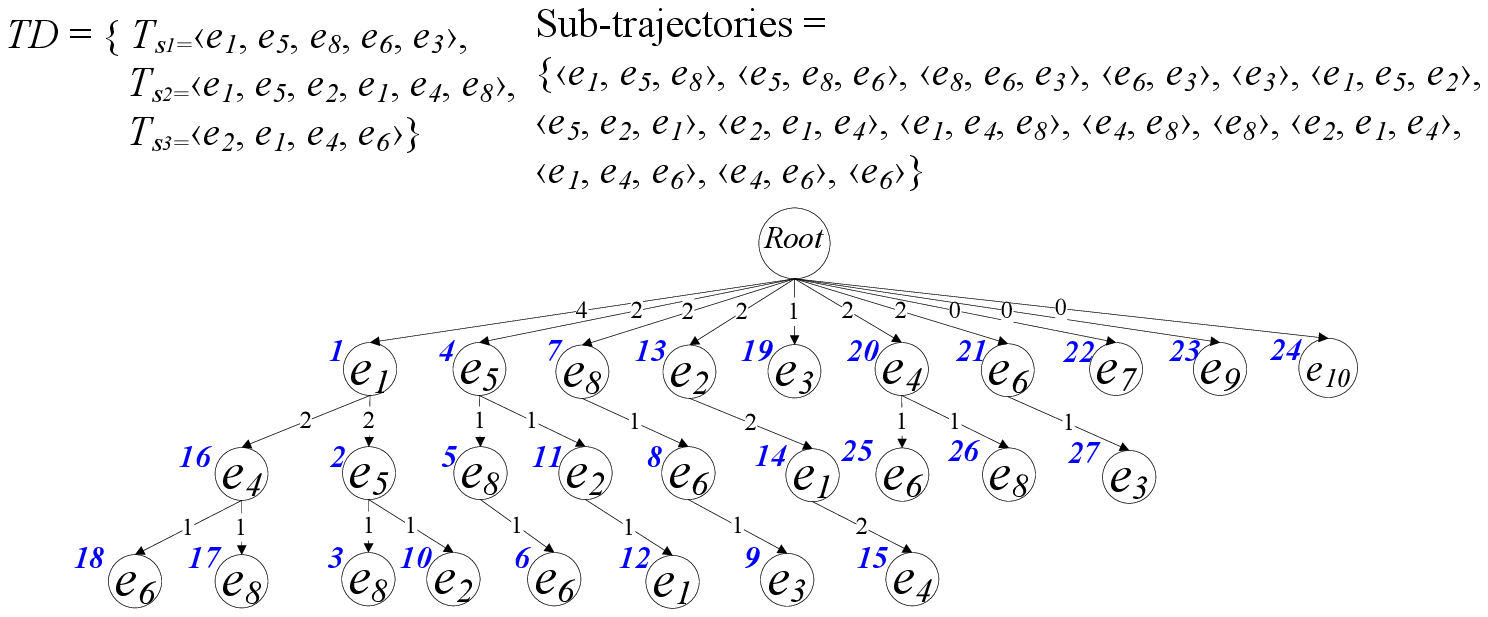,width=0.85\textwidth}
        \vspace{-0.2in}
        \caption{\small Example Trie}
        \label{fig:trie}
\vspace{-0.1in}
\end{figure}

In order to facilitate the understanding of our approach, we assume that an input set $TD$ returned by the first-stage SP compression has three compressed trajectories, i.e., $T_{S_1}$, $T_{S_2}$, and $T_{S_3}$ shown in Fig.~\ref{fig:trie}. Theoretically, we can locate all the sub-trajectories with length ranging from minimum 1 to maximum trajectory length (e.g., ranging from 1 to 6 in our example). However, we set a threshold $\theta$ to only consider the sub-trajectories with their length not exceeding $\theta$. Take the real dataset used in our simulation as an example. On training dataset, by setting $\theta$ as any number from 1 to 20, our approach is able to save around $50\%$ to $70\%$ storage consumption but the time complexity of our approach is proportional to $\theta$. Although we cannot formally find an optimal setting for $\theta$ as it is highly dependent on the training dataset and real trajectory dataset, a small value of $\theta$ already can achieve significant storage saving with reasonable time complexity. In the following discussion, we set $\theta = 3$, which is the optimal length for our trajectory dataset.

For the given input dataset $TD$, we first locate all the sub-trajectories with length not exceeding $\theta$ (i.e., 3). Note we locate one sub-trajectory starting from each edge, so those sub-trajectories near the tail of each trajectory may be shorter than $\theta$. As illustrated in Fig.~\ref{fig:trie}, they are $\langle e_1, e_5, e_8\rangle$, $\langle e_5, e_8, e_6\rangle$, $\langle e_8, e_6, e_3\rangle$, $\langle e_6, e_3\rangle$, $\langle e_3\rangle$, $\langle e_1, e_5, e_2\rangle$, $\langle e_5, e_2, e_1\rangle$, $\langle e_2, e_1, e_4\rangle$, $\langle e_1, e_4, e_8\rangle$, $\langle e_4, e_8\rangle$, $\langle e_8\rangle$, $\langle e_2, e_1, e_4\rangle$, $\langle e_1, e_4, e_6\rangle$, $\langle e_4, e_6\rangle$, and $\langle e_6\rangle$. We then build a Trie based on all identified sub-trajectories. For any node $n$ in the Trie, the path from \emph{root} to $n$ represents a sub-trajectory $T_{sub}$ with the number shown in the link from its parent node indicating the frequency of $T_{sub}$. Here, the number next to each node is the unique ID for the node. Take node 18 as an example. The string formed by the nodes along the path from \emph{root} to node 18 is $ e_1 e_4 e_6$, i.e., corresponding to the sub-trajectory $\langle e_1, e_4, e_6\rangle$. The number 1 shown in the link from node 16 to node 18 represents the frequency of $\langle e_1, e_4, e_6\rangle$, i.e., it only appears once in the training dataset.
%
%
In addition, we want to make sure that the nodes in the first level (the level right below $root$) correspond to all the edges in the original road network. This design is to facilitate the later decomposition process which will be explained later. Take our sample Trie as an example. Assume our original road network consists of 10 edges (i.e., $e_1, e_2, \cdots, e_{10}$), only edges $e_1$, $e_2$, $e_3$, $e_4$, $e_5$, $e_6$, and $e_8$ present as the first edge in the located sub-trajectories. Consequently, we add the rest edges (i.e., $e_7$, $e_9$,and $e_{10}$) to the first level with the corresponding frequency set to zero, as shown in Fig.~\ref{fig:trie}.

\subsubsection{Trajectory decomposition}

\begin{figure}[htb]%
\vspace{-0.12in}
        \centering
        \psfig{file=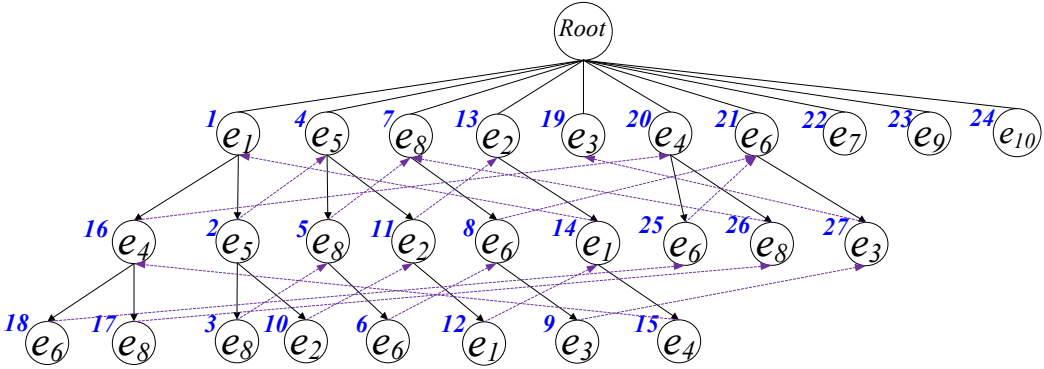,width=0.85\textwidth}
        \vspace{-0.2in}
        \caption{\small Aho-Corasick automaton}
        \label{fig:automaton}
\vspace{-0.1in}
\end{figure}

Once FSTs are identified and the Trie is constructed, we need to decompose an input trajectory into a set of identified FSTs. We borrow the basic idea from Aho-Corasick string matching algorithm. Informally, the algorithm constructs a finite state machine that resembles a trie with additional links between the various internal nodes.
%
%
The automaton is depicted in Fig.~\ref{fig:automaton}, with all the extra links represented by dashed lines. To be more specific, each extra link issued from a node $n_1$ to another node $n_2$ that is the longest possible suffix of the string corresponding to $n_1$. For example, for node 15 ($e_2e_1e_4$), its suffixes are ($e_1e_4$) and ($e_4$). The longest of these that exists in our example is ($e_1e_4$), i.e., node 16. That is why the extra link issued from node 15 points to node 16.

Now we explain how to decompose an input trajectory into a sequence of identified sub-trajectories. It treats the trajectory as a string, and scans the characters (i.e., edges) one by one sequentially. At each step, the current node is extended by finding its children, and if none of the children matches the character, finding its suffix's children, and if that does not work, finding its suffix's suffix's children, and so on, finally ending in the \emph{root} node if nothing has seen before.

\begin{algorithm}[htb]
\begin{small}
\caption{Trajectory decomposition}\label{algo:trajectory-decomposition}
{\bf{Input}}: Aho-Corasick automation $A$, a compressed trajectory $T'=\langle e_1, e_2,\cdots, e_j\rangle$;\\
{\bf{Output}}: a sequence of sub-trajectories;\\
\hspace*{0.03in} {\bf Procedure:}
\begin{algorithmic}[1]
\STATE $n \gets root(A)$; $i\gets 1$; $S\gets \emptyset$; $Res \gets \emptyset$; $l \gets 0$;\\
\WHILE{$i\le |T'| (=j)$}
	\STATE $child \gets Match(n, e_i)$;\\
	\IF{$child \ne -1$}
		\STATE $push(S, child)$; $n\gets child$; $i++$;\\
	\ELSIF{$n.Link_{extra} \ne NULL$}
		\STATE $n\gets n.Link_{extra}$;\\
	\ELSE
		\STATE $n \gets root(A)$;\\
	\ENDIF
\ENDWHILE	
\WHILE{$n \gets pop(S)\ne \emptyset$}
	\IF{$l = 0$}
		\STATE $insert(Res, T_{sub}(n))$, $l \gets |T_{sub}(n)|-1$;\\
	\ELSE
		\STATE $l --$;\\
	\ENDIF
\ENDWHILE
\STATE \textbf{return} $Res$;\\
\end{algorithmic}
\end{small}
\end{algorithm}

Algorithm~\ref{algo:trajectory-decomposition} lists its pseudo code. First, it initializes all the parameters. Here, $n$ indicates the current node of the automaton $A$, $i$ indicates the position of the edge in the trajectory currently evaluated, $S$ is an auxiliary stack holding all the matched nodes in $A$, and $Res$ is the result set which consists of a sequence of the sub-trajectories decomposed from the input trajectory $T'$. It then scans the edges in $T'$ one by one sequentially. For each edge $e_i$, it first checks whether a child of the current node matches $e_i$ via the function $Match(n, e_i)$. If a match occurs, $Match(n, e_i)$ returns the child node. We then push it to $S$, set it as the current node, and proceed to the next edge by increasing $i$ (lines 4-5). If a mismatch occurs indicated by $-1$ returned by $Match(n, e_i)$, we continue the checking at $n$'s suffix node if any via the extra link $Link_{extra}$ (lines 6-7) or the $root$ node (lines 8-9). Recall that for each edge in the original road network, our automaton has a corresponding node in its first level. Consequently, a match can be definitely achieved for each edge in the input trajectory and our decomposition is converged. After the first WHILE-loop (lines 2-9), stack $S$ shall have $|T'|$ (=$j$) nodes, with each corresponding to an edge in $T'$.

Next, we recover the sub-trajectories represented by the nodes in $S$. Note that the initial Aho-Corasick string matching algorithm is a kind of dictionary-matching algorithm that locates elements of a finite set of strings (the ``dictionary") within an input text. As it matches all patterns simultaneously, the returned patterns may have overlaps. However, our purpose for sub-string searching is to decompose the trajectory $T'$ into a sequence of sub-trajectories and hence each edge in $T'$ shall present exactly once in one sub-trajectory. The reason that, when a matched node is found, we do not output the corresponding string but maintain it in the stack $S$ is to avoid the overlapping among different sub-trajectories. As each node in $S$ matches one edge in $T'$, our basic idea is to find the longest matched sub-trajectory from each edge backward. In other words, given a node $n$ with $|T_{sub}(n)|=l$, the next ($l-1$) nodes in $S$ can be ignored. This process is performed by the second WHILE-loop (lines 10-14). Finally, the algorithm returns the sub-trajectories maintained in $Res$ to complete the decomposition. The time complexity of this decomposition process is $O(|T'|)$.

We use an example to illustrate the trajectory decomposition process. Assume $T' = \langle e_1, e_4, e_7, e_5, e_8, e_6, e_3$, $e_1, e_5, e_2$, $e_{10}\rangle$. First, for $e_1$, the first edge in $T'$, it finds a match at node 1, and pushes node 1 to $S$ with $S=\{1\}$. Second, for $e_4$, the second edge in $T'$, it finds a match at node 16 and pushes node 16 to $S$ with $S=\{16, 1\}$. For $e_7$, the third edge of $T'$, it cannot find a match with any child of node 16, and even the child of node 20 (node 16's suffix node). Consequently, we trace-back to the $root$ node and find a match at node 22 and update $S$ to $\{22, 16, 1\}$. The process repeats until all the edges are processed with $S=\{24, 10, 2, 1, 9, 6, 5, 4, 22, 16, 1\}$. Next, we start the sub-trajectory recovery step by popping out nodes from $S$. First, node 24 is popped out, and $T_{sub}(24)=\langle e_{10}\rangle$ is added to $Res$. As $|T_{sub}(24)| = 1$, it does not skip any other node in $S$. Second, node 10 is popped out and $T_{sub}(10)=\langle e_1, e_5, e_{2}\rangle$ is added to $Res$. As its length is three, it skips the next two nodes popped out from $S$, i.e., nodes 2 and 1, but evaluates node 9. It adds $T_{sub}(9)=\langle e_8, e_6, e_{3}\rangle$ to $Res$. Again, the next 2 nodes (i.e., nodes 6 and 5) are skipped and we evaluate node 4 which triggers the insertion of $\langle e_5 \rangle$. This process also continues until $S$ is empty. Finally, $Res = \{\langle e_1, e_4 \rangle,\langle e_7 \rangle, \langle e_5 \rangle, \langle e_8, e_6, e_{3}\rangle, \langle e_1, e_5, e_{2}\rangle, \langle e_{10}\rangle\}$. Accordingly, $T'$ is decomposed into six sub-trajectories, corresponding to nodes 16, 22, 4, 9, 10, and 24, respectively, as shown in Table~\ref{tab:huffman}.

\subsubsection{Encoding procedure}

Finally, we present an encoding procedure which uses Huffman coding to represent the identified FSTs. The main idea is to code each node in Trie. The more frequent a node is, the shorter the code is expected to be. Consequently, we construct a Huffman tree based on all the nodes according to the node frequency, except the \emph{root} node. Huffman tree is a binary tree. A node can be either a leaf node or an internal node. An internal node contains a weight that is a summation of its child nodes' weights, and two links to two child nodes. As a common convention, bit `0' represents following the left child and bit `1' represents following the right child. Assume the initial Trie has $n$ nodes, a corresponding Huffman tree has up to $n$ leaf nodes and $n-1$ internal nodes.

\begin{figure}[htb]%
        \centering
        \psfig{file=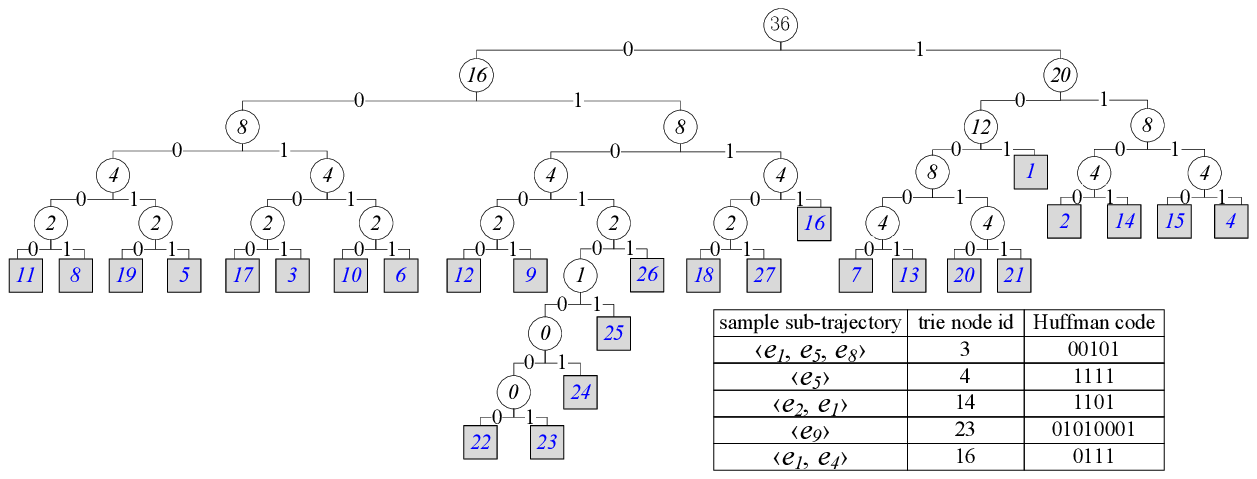,width=0.8\textwidth}
        \vspace{-0.15in}
        \caption{\small Example Huffman tree}
        \label{fig:huffman-tree}
\vspace{-0.15in}
\end{figure}

Initially, all nodes are leaf nodes, and the process essentially begins with the leaf nodes containing the frequencies of the Trie nodes they represent. Then, a new node whose children are the two nodes with smallest frequencies is created, such that the new node's weight is equal to the sum of the children's weight. With the previous two nodes merged into one node, and with the new node being now considered, the procedure is repeated until only one node remains. For the Trie shown in Fig.~\ref{fig:trie}, the corresponding Huffman tree is depicted in Fig.~\ref{fig:huffman-tree}. Here, a rectangle represents a leaf node which corresponds to a node in the Trie, and a circle represents an internal node with the number inside the circle indicating the weight. With the help of Huffman tree, each node of the Trie (i.e., each identified sub-trajectory) can be represented by a unique code. For easy understanding, we list some sample sub-trajectories and their unique codes in Fig.~\ref{fig:huffman-tree}. For example, $\langle e_1, e_5, e_8 \rangle$ is represented by node 3 in the Trie, and its corresponding code is 00101; $\langle e_1, e_4\rangle$ is represented by node 16 in the Trie, and its corresponding code is 0111. Based on Huffman coding, the code for the example trajectory $T'$ is listed in Table~\ref{tab:huffman}.

\vspace{-0.1in}
\begin{table}[!h]
\centering
    \caption{FST compression of trajectory $T'$}
    \label{tab:huffman}
    \begin{small}
    \begin{tabular}{|c|l|}
    \hline
    input $T'$ & $\langle e_1, e_4, e_7, e_5, e_8, e_6, e_3, e_1, e_5, e_2, e_{10}\rangle$\\ \hline
   decomposition & $\langle e_1, e_4 \rangle,\langle e_7 \rangle, \langle e_5 \rangle, \langle e_8, e_6, e_{3}\rangle, \langle e_1, e_5, e_{2}\rangle, \langle e_{10}\rangle$ \\ \hline
    Trie nodes &  16, 22, 4, 9, 10, 24 \\ \hline
    Huffman code& 0111, 01010000, 1111, 01001, 00110, 0101001 \\ \hline
    Result & 011101010000111101001001100101001 \\ \hline
    \end{tabular}
    \end{small}
\vspace{-0.1in}
\end{table}

\subsubsection{Discussion}

As a summary, FST compression first locates all the sub-trajectories with their length not exceeding $\theta$ from the training set, and constructs a Trie. It then builds an Aho-Corasick automaton and a Huffman tree based on the Trie. For a given compressed trajectory $T'$, it decomposes $T'$ into a sequence of sub-trajectories with the help of the automaton, and then uses the Huffman codes of the corresponding sub-trajectories as a compressed format to represent $T'$. The decoding process is straightforward. Given a binary code, it first recovers the sequence of nodes in Trie represented by the binary code with the help of Huffman tree, and then retrieves the sub-trajectories represented by those Trie nodes to recover the trajectory. The time complexity of the first step is in the scale of the length of the binary code. Given the fact that the binary code has $|T|$ as the upper bound, the first step has $O(|T|)$ as the time complexity. The time consumption of the second step is the length of the SP compression result, which is up bounded by $|T|$. Consequently, the time complexity of this step is also $O(|T|)$. Combining two steps, the time complexity of decoding process is $O(|T|)$.

\subsection{Hybrid Spatial Compression (HSC)}


HSC takes advantages of above two spatial compression techniques, and is expected to further improve the compression effectiveness. We assume the all-pair shortest path, the Trie, the automaton and the Huffman tree are constructed in advance. Because the compression and decompression time complexity of both SP compression and FST compression is $O(|T|)$, the compression and decompression time complexity of HSC is $O(|T|)$.



\section{Temporal Compression}
\label{sec:temporal-compression}

We propose to represent the temporal information of a trajectory in the form of $(d_i, t_i)$. This representation is storage consuming, as it suffers from the same scale as the original GPS sampling number. However, on the other hand, each tuple $(d_i, t_i)$ describes when the object is at a specific location. The compression of this information will cause the loss of certain information. Consequently, we propose two metrics to bound the inaccuracy that could be caused by the temporal compression, namely \emph{Time Synchronized Network Distance} (TSND) and \emph{Network Synchronized Time Difference} (NSTD), as formally defined in Definition~\ref{defn:TSND} and Definition~\ref{defn:NSTD}, respectively. To simplify our discussion, we assume the trajectories mentioned in the following are in the format of $((d_1, t_1), (d_2, t_2), \cdots, (d_n, t_n))$. For a given $T$ and a given time stamp $t_x$ ($\in [0, t_n]$), the corresponding distance $d_x$ the object has moved at $t_x$ can be approximated by linear interpolation via function $Dis(T, t_x)$. For example, $Dis(T, t_x)$ with $t_i<t_x\le t_{i+1}$ returns $d_i + \frac{(d_{i+1}-d_i)\times (t_x-t_i)}{(t_{i+1}-t_i)}$. Similarly, for a given $T$ and a given $d_x$ ($\in [0, d_n]$), the corresponding time $t_x$ when object moves $d_x$ distance along $T$ can be approximated by linear interpolation via function $Tim(T, d_x)$.

\subsection{Error metrics}

Before we present our Bounded Temporal Compression (BTC) algorithm, we first introduce the error metrics TSND and NSTD in the following. 

\begin{defn}[Time Syn. Network Dis. (TSND)]\label{defn:TSND}
Given a trajectory $T$ and its compressed one $T'$, TSND measures the maximum difference between the distance object travels via trajectory $T$ and that via trajectory $T'$ at any time slot with $TSND(T, T') = Max_{t_x}(|Dis(T, t_x) - Dis(T', t_x)|)$. $\square$
\end{defn}

\begin{defn}[Network Syn. Time Dif. (NSTD)]\label{defn:NSTD}
NSTD defines the maximum time difference between a trajectory $T$ and its compressed form $T'$ while traveling any same distance with $NSTD(T, T') = Max_{d_x}$ $(|Tim(T, d_x) - Tim(T', d_x)|)$. $\square$
\end{defn}

\begin{figure}[htb]%
\vspace{-0.05in}
        \centering
        \psfig{file=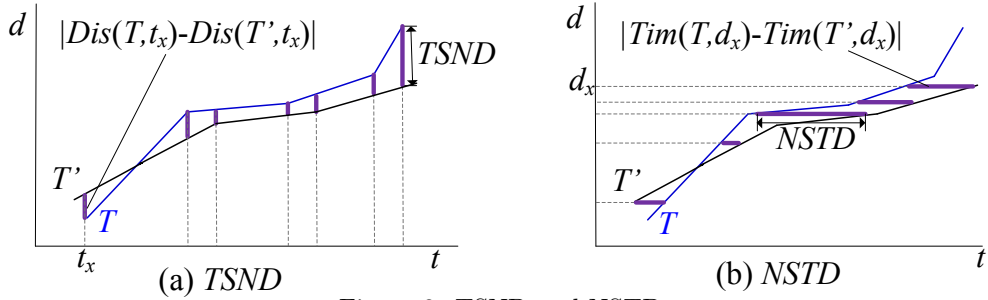,width=0.8\textwidth}
        \vspace{-0.28in}
        \caption{TSND and NSTD}
        \label{fig:tsnd-nstd}
\vspace{-0.1in}
\end{figure}

To facilitate the understanding of these two metrics, we depict an example in Fig.~\ref{fig:tsnd-nstd}. Given a sequence of temporal tuples $T = ((d_1, t_1), (d_2, t_2), \cdots, (d_n, t_n))$, $T$ can be plotted on a $d$-$t$ plane. Then, TSND measures the maximum difference between $T$ and $T'$ along $d$-dimension, and NSTD measures the maximum difference between $T$ and $T'$ along $t$-dimension. We want to highlight that both TSND and NSTD are meaningful only when the compressed trajectory $T'$ keeps exactly the same spatial information as the original trajectory $T'$, which is guaranteed by our HSC algorithm.

The metric TSND is a variant of \emph{Time Synchronized Euclidean Distance} (TSED)~\cite{EDBT04_765,SSDM06_275} metric which measures the distance of two Euclidean space trajectories. Given a Euclidean space trajectory $T_{e}=((l_1, t_1), (l_2, t_2), \cdots, (l_n, t_n))$ and its compressed one $T'_{e}=((l'_1, t'_1), (l'_2, t'_2)$, $\cdots, (l'_m, t'_m))$ with $l_i=(x_i, y_i)$ and $l'_i=(x'_i, y'_i)$ representing two Euclidean points, TSED returns the maximum Euclidean distance between a point $l_i$ and a point $l'_j$ with $(l_i, t_x) \in T_{e}$ and $(l'_j, t_x) \in T'_{e}$, for any $t_x \in [0, t_n]$.

\begin{theo}\label{theo:TSND-lowerbound}
Given an original trajectory $T$ and a compressed trajectory $T'$, if $T'$ is compressed via previously introduced HSC algorithm, $TSND(T, T') \ge TSED(T, T')$.
\end{theo}

\noindent
{\bf{Proof.} }Given $(d_i, t_i) \in T$, let $l_i$ in the form of $(x_i, y_i)$ be the corresponding point that the moving object is located along trajectory $T$ at time $t_i$. Similarly, given $\langle d'_i, t_i \rangle \in T'$, let $l'_i$ in the form of $(x'_i, y'_i)$ be the corresponding point that the moving object is located along trajectory $T'$ at time $t_i$. As $T'$ is compressed via HSC algorithm, $T'$ is exactly the same as $T$ in terms of spatial information although $T'$ takes less space to keep the spatial information than $T$ does. Consequently, $|d_i - d'_i|$ represents the network distance between $l_i$ and $l'_i$. As we know the Euclidean distance between two points is always the lower bound of the corresponding network distance, we have $|d_i - d'_i| \ge dis_{eu}(l_i, l'_i)$. Assume $ TSED(T,T') = dis_{eu}(l_j, l'_j)$ at time $t_j$. As $dis_{eu}(l_j, l'_j) \le |d_j - d'_j|\le TSND(T, T')$, our statement holds and the proof completes. $\blacksquare$

\subsection{Bounded Temporal Compression}

After introducing metrics TSND and NSTD, we are ready to present the \emph{Bounded Temporal Compression (BTC)} algorithm. As $T = ((d_1, t_1), (d_2, t_2), \cdots$, $(d_n, t_n))$ can be plotted as a polygonal line on $d$-$t$ plane, it forms a Euclidean space trajectory in $d$-$t$ space. Consequently, BTC can be transformed to Euclidean trajectory compression, which has been well-studied in the literature. Among available solutions, we adopt an algorithm similar to \emph{Before Opening Window (BOPW)}~\cite{EDBT04_765} because of its excellent performance and the ability to address online trajectory compression issues. The only difference is that the original algorithm purely considers TSED metric, while our implementation considers TSND and NSTD metrics.

The main idea is that for a given trajectory $T$, maximal tolerated TSND $\tau$ and maximal tolerated NSTD $\eta$, BTC scans the tuples in $T$ sequentially. For a tuple $(d_i, t_i)$, it attempts to skip $(d_{i+1}, t_{i+1})$ by linking $(d_i, t_i)$ and $(d_{i+2}, t_{i+2})$ directly. In other words, we attempt to replace the initial sub-trajectory $T_i =((d_i, t_i), (d_{i+1}, t_{i+1}), (d_{i+2}, t_{i+2}))$ with $T'_i =((d_i, t_i), (d_{i+2}, t_{i+2}))$. To evaluate whether this replacement is valid, we calculate NSTD and TSND values between $T_i$ and $T'_i$. If $TSND(T_i, T'_i) \le \tau$ and $NSTD$ $(T_i, T'_i) \le \eta$, this attempt is valid and we can safely skip $(d_{i+1}, t_{i+1})$, and then start the next attempt to skip $(d_{i+2}, t_{i+2})$ by linking $(d_i, t_i)$ and $(d_{i+3}, t_{i+3})$ directly. Otherwise, the attempt is invalid and $(d_{i+1}, t_{i+1})$ cannot be skipped. An invalid attempt terminates the evaluation of tuple $(d_i, t_i)$, and initiates the evaluation of the last successfully attempted tuple $(d_{i+1}, t_{i+1})$. The process repeats until all the edges are evaluated.

\begin{figure}[htb]%
\vspace{-0.13in}
      \centering
      \psfig{file=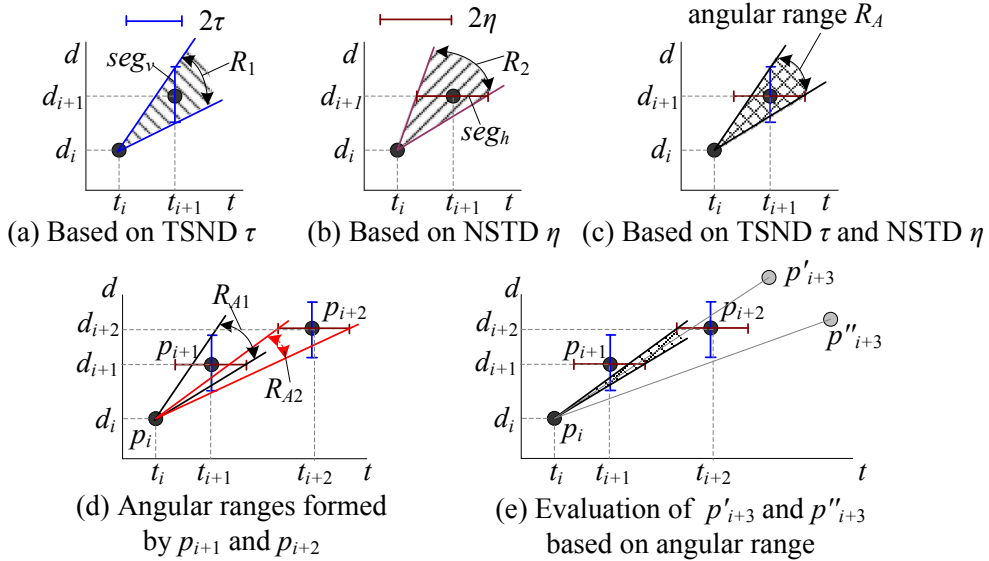,width=0.8\textwidth}
          \vspace{-0.2in}
        \caption{Illustration of angular range}
        \label{fig:angular-range}
\vspace{-0.1in}
\end{figure}

The original implementation of BOPW has a time complexity of $O(|T|^2)$. We improve it to $O(|T|)$ with the help of a novel concept namely \emph{angular range}. Given two points $p_i$ = $(d_i, t_i)$ and $p_{i+1}$ = $(d_{i+1}, t_{i+1})$ in a $d$-$t$ space, we assume BOPW keeps $p_i$ in $T'$. No matter how $T'$ looks like, it must satisfy $\tau$ and $\eta$, i.e., $TSND(T,T') \le \tau$ and $NSTD(T,T') \le \eta$. In other words, the difference between $T$ and $T'$ along $d$ dimension at $t_{i+1}$ is bounded by $\tau$ and the difference between $T$ and $T'$ along $t$ dimension at $d_{i+1}$ is bounded by $\eta$. Consequently, given a vertical line segment $seg_v$ centered at $d_{i+1}$ with $|seg_v| = 2 \tau$, $T'$ must intersect $seg_v$. As shown in Fig.~\ref{fig:angular-range}(a), $seg_v$ bounds an angular range $R_1$ that $T'$ shall fall within. Similarly, given a horizontal line segment $seg_h$ centered at $t_{i+1}$ with $|seg_h|=2 \eta$, $T'$ must intersect $seg_h$. As shown in Fig.~\ref{fig:angular-range}(b), $seg_h$ actually bounds an angular range $R_2$ that $T'$ shall fall within. Considering both $\eta$ and $\tau$, the angular range is shrunk to the intersection between $R_1$ and $R_2$, i.e., the shaded angular range $R_A$ depicted in Fig.~\ref{fig:angular-range}(c).

In order to facilitate the presentation, we assume function $R_A(p_i, S_p, \tau, \eta)$ returns the angular range centered at point $p_i$ formed by all the points of set $S_p$, i.e., $R_A(p_i, S_p, \tau, \eta) = \cap_{\forall p\in S_p}R_A(p_i, \{p\}, \tau, \eta)$. Take points depicted in Fig.~\ref{fig:angular-range}(d) as an example, $R_A(p_i, \{p_{i+1}\}, \tau, \eta)$ = $R_{A_1}$, $R_A(p_i, \{p_{i+2}\}, \tau, \eta)$ = $R_{A_2}$, and $R_A(p_i, \{p_{i+1}, p_{i+2}\}, \tau, \eta)$ = $R_{A_1}\cap R_{A_2}$.

\begin{algorithm}[htb]
\begin{small}
\caption{Bounded Temporal Compression}\label{algo:tem-compression}
{\bf{Input}}: a trajectory $T=((d_1,t_1), (d_2,t_2), \cdots, (d_n,t_n))$;\\
{\bf{Output}}: a compressed trajectory $T'$;\\
\hspace*{0.03in} {\bf Procedure:}
\begin{algorithmic}[1]
\STATE $index \gets 1$, $T'\gets p_{index}$, $R\gets [-\frac{\pi}{2},\frac{\pi}{2}]$; \\
\FOR{$i\gets 2$ to $n$}
    \IF{$FallInside(R, p_{index}, p_i) = 1$}
        \STATE $R \gets R \cap R_A(p_{index}, \{p_{i}\}, \tau, \eta)$;\\
    \ELSE
        \STATE $Append(T', p_{i-1})$; $R\gets [-\frac{\pi}{2},\frac{\pi}{2}]$; $index \gets i-1$;\\
    \ENDIF
\ENDFOR
\STATE \textbf{return} $T'$;\\
\end{algorithmic}
\end{small}
\end{algorithm}


With the help of angular range, we can compress $T = \langle p_1, p_2, \cdots, p_n \rangle$ based on BOPW with $O(|T|)$ time complexity, and its pseudo code is listed in Algorithm~\ref{algo:tem-compression}. It maintains a pointer $index$ pointing to the last point $p_{index}$ along $T$ that has been enrolled into $T'$, and an angular range $R$ centered at $p_{index}$ that bounds all the possible rotations $T'$ can make right after $p_{index}$. Initially, $R$ is set to straight angle $[-\frac{\pi}{2},\frac{\pi}{2}]$ which captures the full half-plane after $p_{index}$. Thereafter, $R$ gets shrunk by the points evaluated. For each point $p_i$ that is scanned by the algorithm, we check whether $p_i$ is located inside the current angular range $R$ centered at $p_{index}$ via the boolean function $FallInside(R, p_{index}, p_i)$ and there are two possible outputs. If $p_i$ is within $R$, the points $p_{index+1},p_{index+2}, \cdots, p_{i-1}$ can be skipped. We further shrink the current range $R$ by $p_i$ and start the evaluation on $p_{i+1}$. Otherwise, $p_{i-1}$ cannot be skipped and we append $p_{i-1}$ to $T'$. Meanwhile, a new angular range centered at $p_{i-1}$ is initiated. Take Fig.~\ref{fig:angular-range}(e) as an example. Suppose $index = i$ and currently we are evaluating point $p_{i+3}$. If $p_{i+3}$ is located at the point of $p'_{i+3}$, $p_{i+3}$ is within the angular range $R$ (i.e., the shaded area). We can shrink $R$ based on $R_A(p_i, \{p_{i+3}\}, \tau, \eta)$ and then continue the process. If $p_{i+3}$ is located at the point of $p''_{i+3}$, it is located outside the current $R$. Consequently,
the algorithm will enroll $p_{i+2}$ into $T'$, set $index$ to $(i+2)$ and a new angular range $R$ to straight angle, and then continue the process.


\section{Applications on Compressed Trajectory}
\label{sec:application}

The main purpose of trajectory compression is to use less space to store the trajectories. Consequently, whether the compressed trajectories can support various LBS applications is \emph{not} the main focus of different compression approaches. As PRESS compresses the trajectories in such a way that the spatial paths are captured exactly and the temporal information loss is bounded by TSND and NSTD, we can decompress the trajectories for LBS applications. However, it is still desirable that the compressed trajectories can support certain, if not all, applications without being fully decompressed. In the following, we demonstrate in detail that the compressed trajectory can support $where_{at}$, $when_{at}$ and $range$, three common queries used by many LBSs, and briefly introduce some other queries PRESS can support. $where_{at}(T, t)$ returns a location along the trajectory $T$ where an object is located at time $t$, $when_{at}(T,x,y)$ returns a time stamp when an object is located at $(x,y)$ while traveling along $T$, and $range(T,t_1,t_2,R)$ checks whether trajectory $T$ passes the region $R$ during time period $t_1$ to $t_2$.

\subsection{$where_{at}$ Query}

$where_{at}(T, t)$ returns a location along the trajectory $T$ where an object is located at time $t$. Given a trajectory $T$, its compressed form $T'$ returned by PRESS, and a time slot $t_i$, let $Dis(T, t_i)=d_i$, $Dis(T',t_i)=d'_i$, $where_{at}(T, t_i)=p_i$ and $where_{at}(T', t_i)=p'_i$. Based on the fact that $|d_i - d'_i| \le TSND(T,T')$ guaranteed by BTC, we have $|where_{at}(T', t_i)-where_{at}(T, t_i)|=|p_i-p'_i|\le TSND(T,T')$. This is because $|p_i-p'_i|$ refers to the shortest distance from $p_i$ to $p'_i$, and $|d_i - d'_i|$ refers to the distance from $p_i$ to $p'_i$ along trajectory $T$. Obviously, $|p_i-p'_i|\le|d_i - d'_i|\le TSND(T,T')$.

Given one original trajectory, we assume its spatial information is represented by $n$ edges $\langle e_1, e_2, \cdots, e_n \rangle$ and its temporal information is captured by $m$ $(d_i, t_i)$ tuples. Query $where_{at}(T, t)$ needs to first locate $d$ based on $t$ (i.e., $Dis(T,t)$ $=d$) and then locate $e$ based on $d$. Consequently, it visits $\frac{m}{2}$ temporal tuples and $\frac{n}{2}$ edges on average.

Given a compressed trajectory $T'$, $where_{at}(T', t)$ query still needs to locate $d'$ based on $t$. Assume the temporal compression ratio is $\frac{|T|}{|T'|} = \beta$, it needs to scan $\frac{m}{2\beta}$ tuples on average. Thereafter, it needs to locate the point along $T'$ corresponding to the distance $d'$. As this process relies on some additional information, we first introduce the auxiliary structures. We assume certain distance information is embedded in Trie to facilitate the query processing. For each node $n$ in Trie, it stores the distance of $T_{sub}(n)$, denoted as $T_{sub}(n).d$. For example, node 16 keeps the distance of $T_{sub}(16)=\langle e_1, e_4\rangle$, and node 9 keeps the distance of $T_{sub}(9)=\langle e_8, e_6, e_3 \rangle$. Note that the sub-trajectories captured by Trie might not be a real sub-trajectory of any original trajectories, as Trie takes all the compressed trajectories of SP compression as input. Consequently, we need to decompress the sub-trajectory $T_{sub}(n)$ based on SP decompression in order to calculate the distance $T_{sub}(n).d$. In addition to this node distance, we also assume the distance of all-pair shortest paths is maintained by the shortest path table.

Now, we are ready to explain how to locate the answer point along the trajectory based on a given $d'$. Given a binary code, we maintain an accumulative distance $d_{acu}$ and recover the sequence nodes $n_i$ one by one following the FST decompression process. For each recovered node $n_i$, we increase $d_{acu}$ by $T_{sub}(n_i).d$. In addition, we need to check whether the hop from $T_{sub}(n_{i-1})$ to $T_{sub}(n_i)$ is seamless. To achieve this, we get the character (i.e., edge) represented by $n_{i-1}$ (i.e., the node right before $n_i$), that is the last edge of $T_{sub}(n_{i-1})$; and we get the first character along the path from $root$ to $n_i$, that is the first edge of $T_{sub}(n_i)$. We then check the shortest-path table to get the shortest distance between them, which also contributes to $d_{acu}$. After recovering a node $n_i$, we check whether $d_{acu} < d'$. If yes, the process continues to recover the next node $n_{i+1}$; otherwise, the answer point must locate in $T_{sub}(n_i)$. Assume $T_{sub}(n_i)$ is in the form of $\langle e_j, e_{j+1}, \cdots, e_{j+u}\rangle$. We then scan the edge and its immediate follower one by one. Initially, we check $e_j$ and $e_{j+1}$, get their shortest distance from the distance table, and add it to $d_{acu}$. If $d_{acu}<d'$, it proceeds to $e_{j+1}$ and $e_{j+2}$. Otherwise, the answer point must locate at the shortest distance from $e_j$ to $e_{j+1}$. We then fully recover the shortest path and find the answer point. Assume that the SP compression ratio is $\alpha$, and the FST compression ratio is $\gamma$. It on average recovers $\frac{n}{2\alpha\gamma}$ Trie nodes, and checks $\frac{\gamma}{2}$ edges within the located sub-trajectory. Given the fact that $\alpha > 1$ and $\gamma>1$, the time complexity is reduced.

\subsection{$when_{at}$ Query}

Before introducing $when_{at}$, we show that $when_{at}$ is error-bounded by $NSTD$, i.e., for a trajectory $T$ and its compressed form $T'$ compressed via PRESS, $|when_{at}$ $(T,x,y) - when_{at}(T',x,y)|=|t_i-t'_i|\le NSTD(T,T')$. 
Given an input point $p =(x,y)$ and $T$, $when_{at}$ assumes $p$ is a point along $T$ and locates $p$ to an edge of $T$. Then, it derives the network distance $d$ traveled along $T$ from the starting point until $p$, based on which the corresponding time $t_i$ can be located along the temporal information of $T$. As HSC is errorfree, $T'$ captures the exact spatial information as $T$ and the corresponding $d'$ derived based on $T'$ and $(x,y)$ equals $d$. As guaranteed by Definition~\ref{defn:NSTD}, $|t_i-t'_i| \le NSTD(T,T')$.

Similar as $where_{at}(T, t)$, $when_{at}(T, x, y)$ needs to first locate $(x,y)$ to an edge $e_j \in T$ along the trajectory such that $(x,y) \in e_j$, derive the distance $d_i$ traveled along $T$ until the input point $(x,y)$, and then locate $t_i$ along the temporal representation based on $d_i$. On average, it visits $\frac{n}{2}$ edges and $\frac{m}{2}$ temporal tuples. Given a compressed trajectory $T'$, $when_{at}(T, x, y)$ is processed similarly except it actually visits fewer edges. To facilitate the process, we record an MBR (Minimum Bounding Rectangle) for each Trie node $n$ and an MBR for the shortest path of each pair of nodes. When we recover the nodes one by one, we check whether $(x,y) \in MBR(n)$. If $(x,y) \in MBR(n)$, we scan $T_{sub}(n)$ edge by edge and check whether $(x,y) \in MBR(e_i, e_j)$, assuming $e_i$ and $e_j$ are two adjacent edges in $T_{sub}(n)$. Every time when $(x,y) \in MBR(SP(e_i, e_j))$, we retrieve the shortest path $SP(e_i, e_j)$ and map $(x,y)$ back to $SP(e_i, e_j)$. This process might be repeated a few times as MBRs are much bigger than $SP(e_i, e_j)$ and the fact $(x,y) \in MBR(SP(e_i, e_j))$ does not guarantee $(x,y) \in SP(e_i, e_j)$. Once we get the distance, we can locate the time based on compressed temporal sequence. As a summary, $when_{at}(T, x, y)$ spends $\frac{n}{2\alpha\gamma}+\frac{\alpha+\gamma}{2}$ time unit on spatial process and $\frac{m}{2\beta}$ time unit on temporal process.

\subsection{$range$ Query}

For boolean range query on an original trajectory $T$, it first locates $d_1$ and $d_2$ based on $t_1$ and $t_2$ on the temporal sequences, and then retrieves the spatial segment $Seg$ between $d_1$ and $d_2$. It then scans $Seg$ edge by edge and checks whether any edge intersects the query region $R$. The process on a compressed trajectory $T'$ is similar. Given $t_1$ and $t_2$, it locates $d'_1$ and $d'_2$ as described in $when_{at}$ query, and then retrieves points corresponding to $d'_1$ and $d'_2$. As we maintain the MBRs for all the shortest paths and the sub-trajectories captured by Trie, we can first check whether an MBR overlaps with $R$ before recovering the original sub-trajectories. Its time complexity is the same as that under $when_{at}$ query.
%
%

\subsection{Discussion}

For all the three queries studied above, the processing over compressed trajectories demonstrates certain non-negligible advantages, compared with the processing over original trajectories. We agree that the gain in terms of performance has a cost of enlarged storage cost, e.g., the storage cost for maintaining the distance for all-pair shortest paths and that for maintaining the MBRs for all-pair shortest paths with both in the scale of $|V|^2$. However, all these auxiliary structures can be pre-processed and can be used for a relatively long duration unless the road network structure changes and/or the movement patterns of the underlying trajectories change significantly. Compared with the large number of trajectories generated daily and the long time period of collection we have to maintain, the extra storage cost incurred by these auxiliary structures can be well-justified. We will further demonstrate it in our simulation study to be presented in Section~\ref{sec:exp}.

PRESS also supports other queries commonly used by LBSs. For instance, it can support queries inquiring trajectories passing near a location point $(x, y)$ within distance $d$ from $t_1$ to $t_2$ and queries calculating the minimal distance between two trajectories $T_1$ and $T_2$.

For the query that looks for trajectories passing near a location point $(x, y)$ within distance $d$ during a time period $t_1$ to $t_2$, the process is similar to that of $range$ query. For each trajectory $T$ in dataset $TD$, it first invokes $where_{at}(T, t_1)$ and $where_{at}(T, t_2)$ to get the distance range in temporal sequence $d_1$ to $d_2$, and then retrieves the spatial segment $Seg$ between $d_1$ and $d_2$. We can skip a whole encoded sub-trajectory or shortest path if its minimum distance to $(x, y)$ is longer than $d$. Only the undetermined segments will be de-compressed and further checked. At last, we will decide whether the trajectory passes near the specific location point. The acceleration happens during the $where_{at}$ query processing and the skipping strategy we use during the judgment process.

For the query calculating the minimal distance between two trajectories $T_1$ and $T_2$, the original approach may be calculating the distance of each pair of edges the two trajectories pass and return the one with minimum distance. By maintaining the MBRs of Trie-denoted sub-trajectories and shortest paths, however, we can skip the comparison inside two FST segments if the minimal distance of their MBRs exceeds our already obtained answer, or skip the comparison inside two SP segments similarly. As a result, the time spent on the comparison is much shorter than the original one.

We believe that based on the basic queries presented in the paper and other potential ones we do not mention, PRESS is able to support many advanced LBSs. We list three applications as examples in the following.
\begin{itemize}
\item By using the $where_{at}$ query on all trajectories at a specific time $t$ in one particular day, we can get the snapshot of the traffic condition of time $t$, based on which further traffic analysis could be performed. For instance, we can use clustering approaches to find out congested regions.

\item By using $range$ query or passing nearby query we illustrated above, we can get the traffic flow of specific regions.

\item By combining $range$ query with minimal distance query of trajectories, we can estimate the similarity of trajectories and further mine behavior patterns behind the generator of such trajectories.
\end{itemize}

On the other hand, we have to admit that PRESS cannot support all queries. This is because after transforming the trajectories into another format, we have to sacrifice some former properties before the compression. Such shifting of properties is sure to bring some drawbacks (becoming unable to support some queries) along with the advantages they take, e.g., accelerating some other queries. Take the "most commonly travelled path" query as an example. In order to get an "edge-level" precise result using a statistical approach (counting the frequency of each edge travelled by all trajectories), we have to first decompress the compressed spatial path into edge sequence and then conduct the counting process. However, this does not counteract the benefits PRESS brings to trajectory. Taking the high compression ratio and those queries we support into consideration, the sacrifice has been well paid off. We would like to give an analogy between PRESS and JPEG, the most famous compression approach in digital image processing. The result of JPEG compression is a pixel image supporting no advanced queries, even basic editing operations. Nevertheless, we still consider JPEG as a classical and valuable approach because of the enormous contribution it brings to saving the storage cost of images. Now the majority of internet transferred images are in JEPG format. Similarly, PRESS, thanks to its superior compression power, lossless nature of the spatial information, and error-bounded nature of the temporal information, has a potentially big market.
%

\section{Experiments Study}
\label{sec:exp}

In this section, we conduct extensive experiments to demonstrate the effectiveness and efficiency of PRESS. The experiments are based on real trajectory data from one of the largest taxi companies in Singapore. Each taxi has installed GPS, and it reports its locations regularly. In our studies, we use the trajectories reported within January 2011, in total $465,000$ trajectories generated by about 15,000 taxis. The original storage cost of this dataset is 13.2GB. First, we map the GPS locations using the approach proposed in~\cite{GIS12_605} to get the spatial path of the trajectories. Then, we project the sample points onto the spatial path and calculate the distance from the starting point of the trajectory by linear interpolation to generate the temporal presentation of the trajectories. Our source code is available at https://github.com/RenchuSong/PRESS.

In addition to PRESS framework, we implement \emph{Map-matched trajectory compression} (MMTC)~\cite{JSS13} and \emph{Nonmaterial}~\cite{ICDT05_173}, two state-of-the-art approaches for trajectory compression in road networks, as the representatives of existing approaches. All the algorithms are implemented with C/C++ and run on a computer with Intel Core i7-3770 CPU (3.40 GHz) and 32 GB memory. In the following, we first evaluate the effectiveness of various approaches by reporting the compression ratio; next evaluate the efficiency of various approaches by reporting the time taken; and finally report the flexibility of various approaches by demonstrating their capabilities in supporting common LBSs.

\subsection{Compression Effectiveness}

In the first set of experiments, we first discuss the effectiveness of our spatial compression algorithm HSC and our temporal compression algorithm BTC, and then compare the compression effectiveness of various algorithms. We adopt \emph{compression ratio} as the performance metric. Given a trajectory $T$ and its compressed form $T'$, the compression ratio is defined as the ratio of $T$'s storage cost to $T'$'s storage cost, i.e., $\frac{|T|}{|T'|}$. 

\begin{figure}[htb]%
\vspace{-0.1in}
\begin{center}
\subfigure[SP Compression]{
   \label{fig:cr-HSTC1}
   \psfig{file=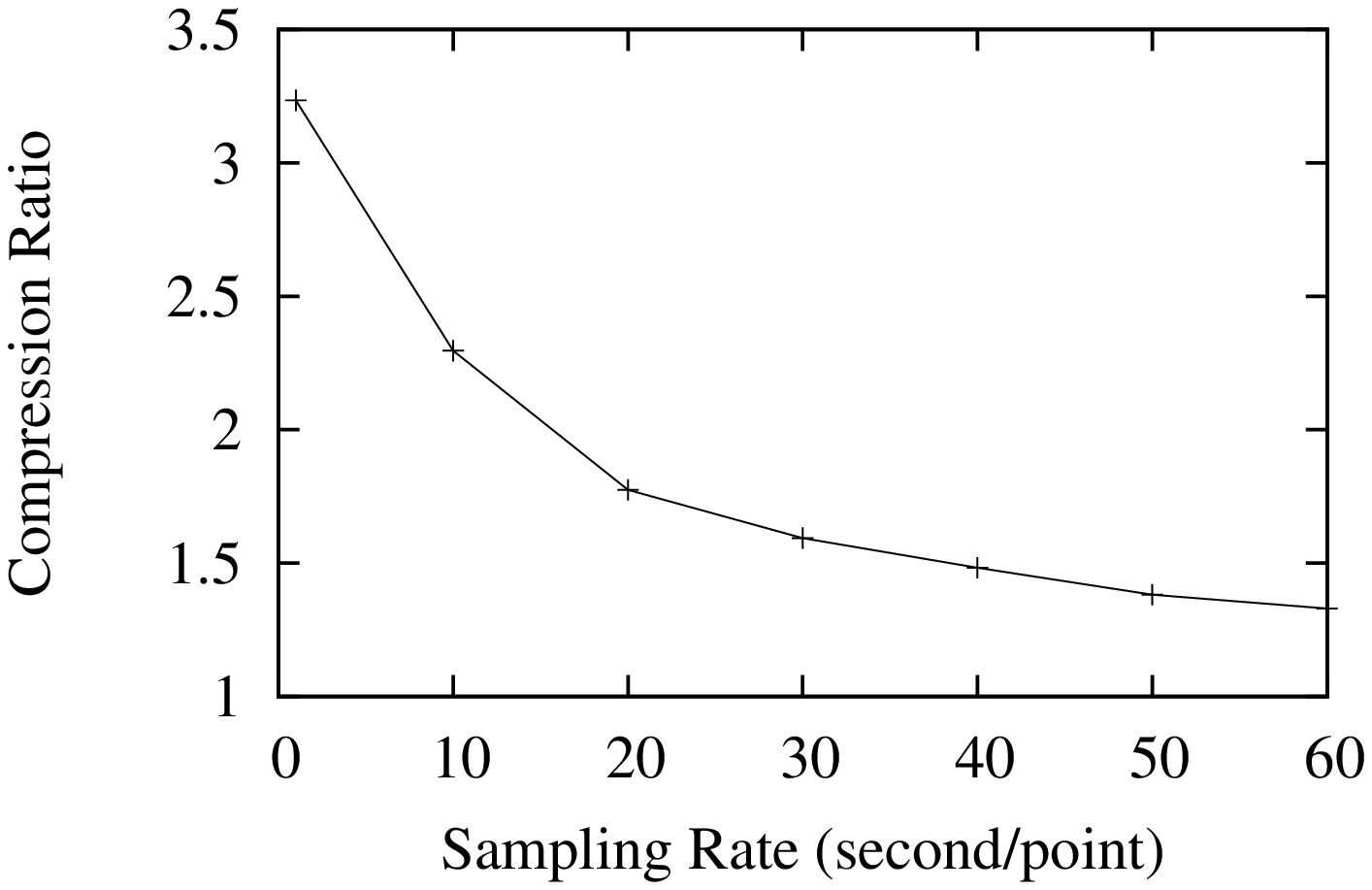,angle=270,width=0.4\textwidth}
   \vspace{-0.1in}
  }
\hspace{-0.1in}
\subfigure[FST Compression]{
   \label{fig:cr-HSTC2}
   \psfig{file=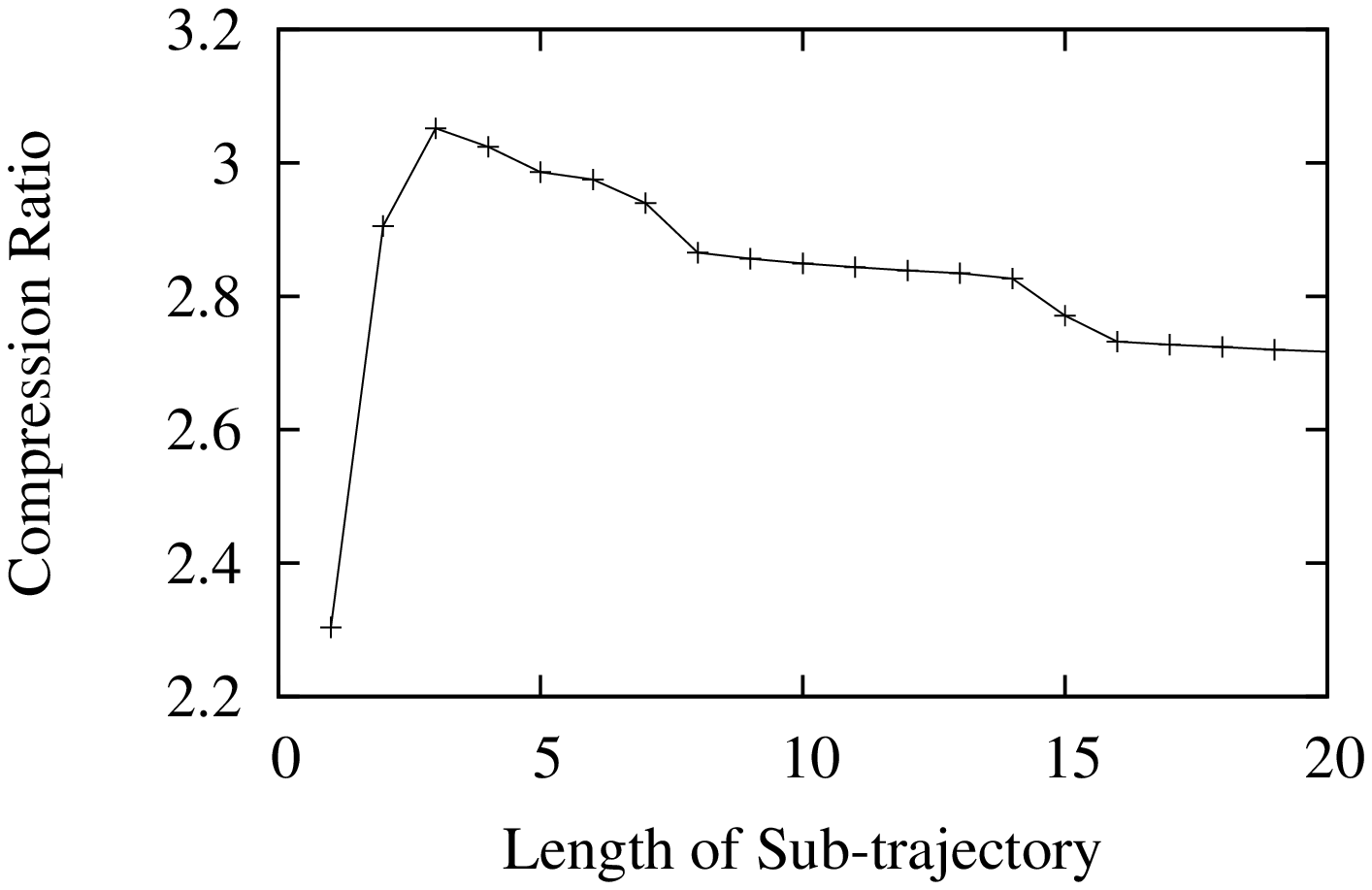,angle=270,width=0.4\textwidth}
   \vspace{-0.1in}
  }
\vspace{-0.1in}
\caption{Compression Ratio of HSC}
\label{fig:cr-HSTC}
\end{center}
\vspace{-0.28in}
\end{figure}

First, we report the effectiveness of the spatial compression algorithm HSC. As introduced in Section~\ref{sec:spatial-compression}, HSC is a two-stage process, compressing the spatial path based on SP compression first and then based on FST compression. Given a spatial path in a road network, the storage cost of original trajectory $T$ relies on the sampling rate. The higher the sampling rate, the more the points $T$ contains with higher storage cost. However, the sampling rate does not affect SP compression that much.
%
As shown in Fig.~\ref{fig:cr-HSTC1}, we demonstrate the power of SP compression under different sampling rate. Via changing the sampling rate from 1 second/point to 60 seconds/point, the SP compression on average can achieve a compression ratio of 1.52. It is very close to the compression ratio of the sampling rate 30 seconds/point, which is also the median sampling rate of the trajectory dataset we use.

\begin{figure}[htb]%
\vspace{-0.1in}
\begin{center}
\subfigure[Compression Ratio]{
   \label{fig:fst-cr}
   \psfig{file=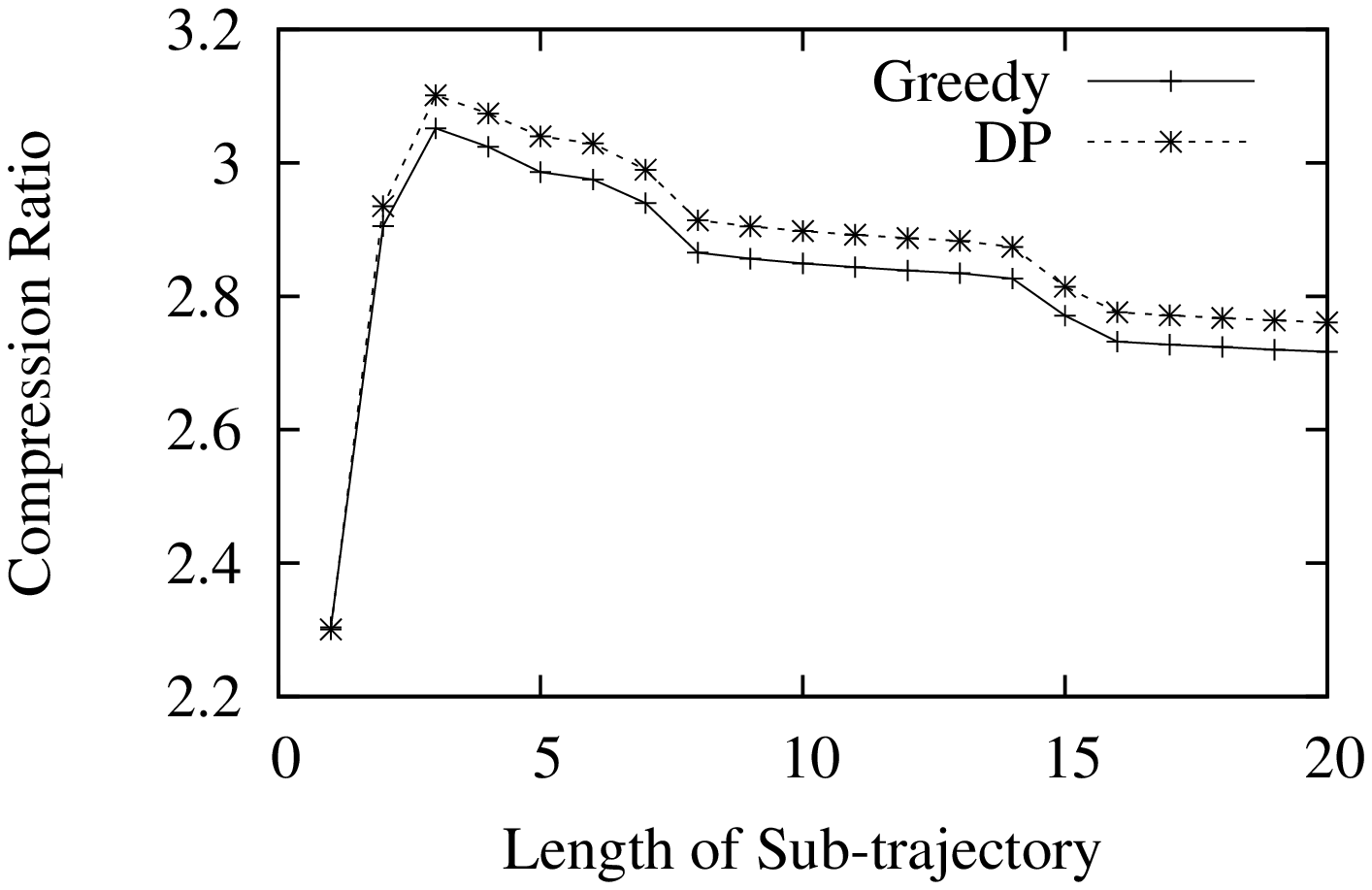,angle=270,width=0.4\textwidth}
   \vspace{-0.1in}
  }
\hspace{-0.1in}
\subfigure[Time Efficiency]{
   \label{fig:fst-time}
   \psfig{file=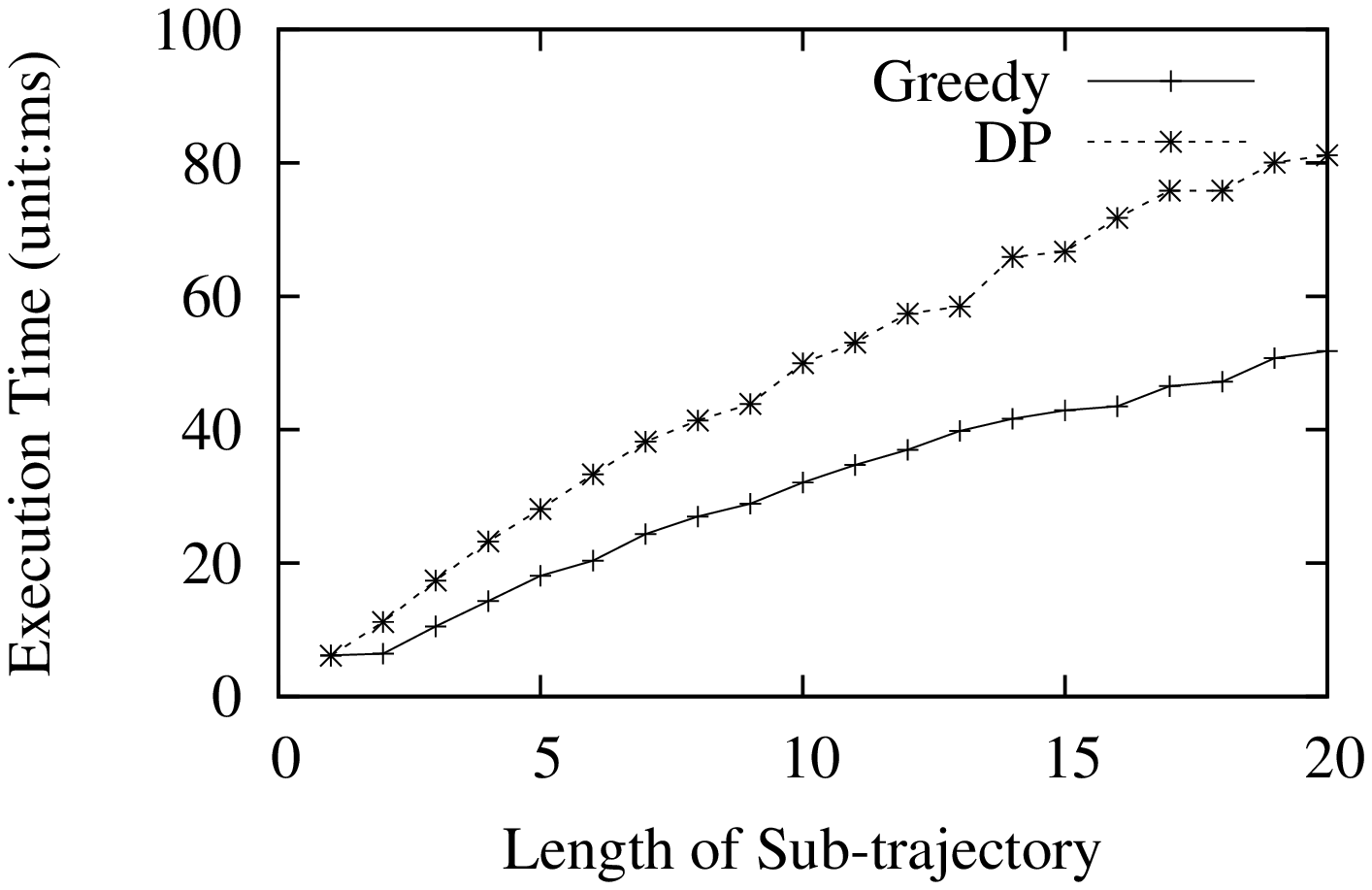,angle=270,width=0.4\textwidth}
   \vspace{-0.1in}
  }
\vspace{-0.1in}
\caption{Performance of FST Compression}
\label{fig:fst-perfomance}
\end{center}
\vspace{-0.2in}
\end{figure}

In the second stage, we compress the spatial path via FST, with its compression ratio under different $\theta$ (i.e., the maximum length of the sub-trajectories in Trie) reported in Fig.~\ref{fig:cr-HSTC2}. Note, assume $T$ is an original spatial path, $T'$ is the compressed trajectory via SP compression, and $T''$ is the compressed trajectory via FST compression, the compression ratio reported here is the ratio of $T''$'s storage cost to $T'$'s (but not $T$'s) storage cost. We use the trajectory set corresponding to one day as the training set, and change the $\theta$ values from 1 to 20. As shown in Fig.~\ref{fig:fst-perfomance}(a), the compression ratio improves initially with the increase of $\theta$. However, when $\theta >3$, a bigger $\theta$ does not help improve the compression ratio. This is because a larger $\theta$ corresponds to a Trie with more nodes, and hence the Huffman tree has to use more bits to represent each node. We set $\theta = 3$ in the following experiments. In real compression applications, the optimal value of $\theta$ can be obtained by attempting to compress a subset of the complete trajectory dataset. Our FST compression can achieve the optimal compression ratio of 3.05 when $\theta =3$. In other words, by combining these two stages (i.e., SP compression and FST compression), our HSC is expected to have an optimal compression ratio of $1.52\times 3.05 \approx 4.64$.

Recall that for FST compression, we propose a greedy algorithm to decompose a given trajectory into a sequence of sub-trajectories. However, there is actually an optimal approach that results in the maximum compression ratio based on dynamic programming. Given a trajectory $T'$ compressed via SP compression, the main idea is to invoke dynamic programming to calculate the splitting solution of $T'$. Assume $T' = \langle e_1, e_2, \cdots, e_i\rangle$ and $F_k$ is the minimum storage cost of the prefix $k$ edges of $T'$, then $F_k = min_{\forall j<k}(F_j + Huf(e_{j+1}e_{j+2}$ $\cdots e_{k}))$. Here, $Huf(S_{jk})$ represents the Huffman code length of a string $S_{jk}$ in the Trie. Initially, $F_0 = 0$. Thereafter, we can derive the optimal splitting point based on dynamic programming. Although this algorithm can generate a compressed trajectory with minimum storage cost, its time complexity is high. Alternatively, we propose a greedy algorithm, as presented in Section~\ref{sec:spatial-compression}. In order to demonstrate that our greedy algorithm can achieve a performance comparable with that of dynamic programming but with a much cheaper time cost, we compare these two approaches in Fig.~\ref{fig:fst-perfomance}. As observed from Fig.~\ref{fig:fst-perfomance}(a), the compression ratio of our greedy algorithm is almost the same as that achieved by dynamic programming, with only around $1\%$ difference. However, as shown in Fig.~\ref{fig:fst-perfomance}(b), our greedy algorithm is much faster, as it only incurs around $65\%$ of dynamic programming's time on average. Consequently, our choice of greedy algorithm is well justified.

\begin{figure}[htb]%
\vspace{0in}
\begin{center}
\subfigure[BTC]{
   \label{fig:compress-temp}
   \psfig{file=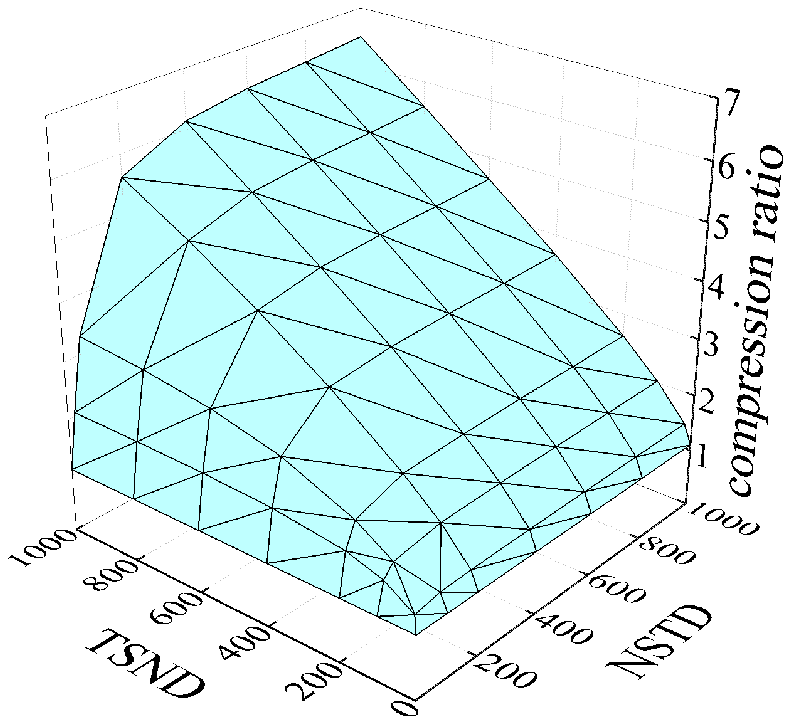,width=0.4\textwidth, height=0.3\textwidth}
   \vspace{-0.1in}
  }
\hspace{0.15in}
\subfigure[PRESS]{
   \label{fig:compress-both}
   \psfig{file=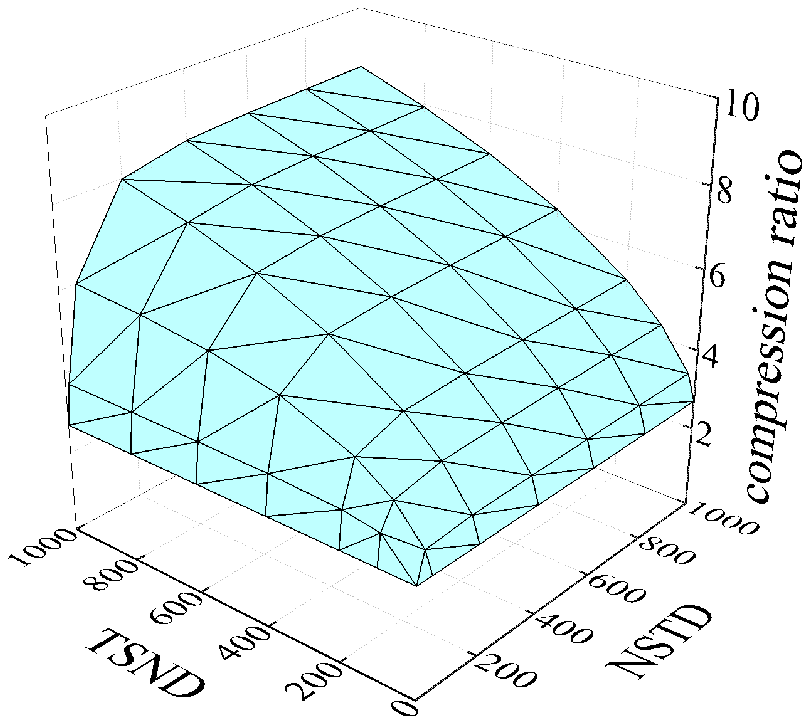,width=0.4\textwidth,height=0.3\textwidth}
   \vspace{-0.1in}
  }
\vspace{-0.15in}
\caption{Compression Ratio vs. TSND and NSTD}
\end{center}
\vspace{-0.2in}
\end{figure}

After presenting the performance of spatial compression, we are ready to present the performance of temporal compression. As our temporal compression algorithm BTC takes TSND and NSTD as inputs, we report the compression ratio under different TSND and NSTD settings in Fig.~\ref{fig:compress-temp}. To be more specific, the value of TSND changes from 0, 10, 20, 50, 100, 200, 400, 600, 800, to 1000 (unit: meter), and the value of NSTD changes from 0, 10, 20, 50, 100, 200, 400, 600, 800, to 1000 (unit: second). It is noticed that even when TSND = NSTD = 0, BTC can still achieve a compression ratio of 1.1. This is because around $10\%$ of the trajectory sampling points tell that the taxi is not moving (e.g., waiting at a taxi drop-off point, or stuck in a traffic jam). In general, the larger the TSND and/or NSTD are, the higher the compression ratio is, which is consistent with our expectation. For example, when TSND = 1000m and NSTD = 1000s, the compression ratio is as high as 6.49.

Now considering both spatial compression and temporal compression, we plot the overall compression ratio of our newly compression framework PRESS in Fig.~\ref{fig:compress-both}, under various TSND and NSTD values. We assume the initial trajectory is represented by $(x,y,t)$ tuples. We then re-format the trajectory as a spatial path and a time sequence, and invoke HSC to compress the spatial path and invoke BTC to compress the time sequence based on TSND and NSTD settings. Our approach is effective. Even when TSND = NSTD = 0, PRESS still achieves a compression ratio of 2.71 (i.e., can save around $63\%$ storage cost). As TSND and/or NSTD increase, the compression ratio is also improved. For example, when TSND = 1000m and NSTD = 1000s, the compression ratio is 8.52.

%

\begin{figure}[htb]
\centering
    \begin{minipage}[t]{.4\textwidth}
        \centering
        \psfig{figure=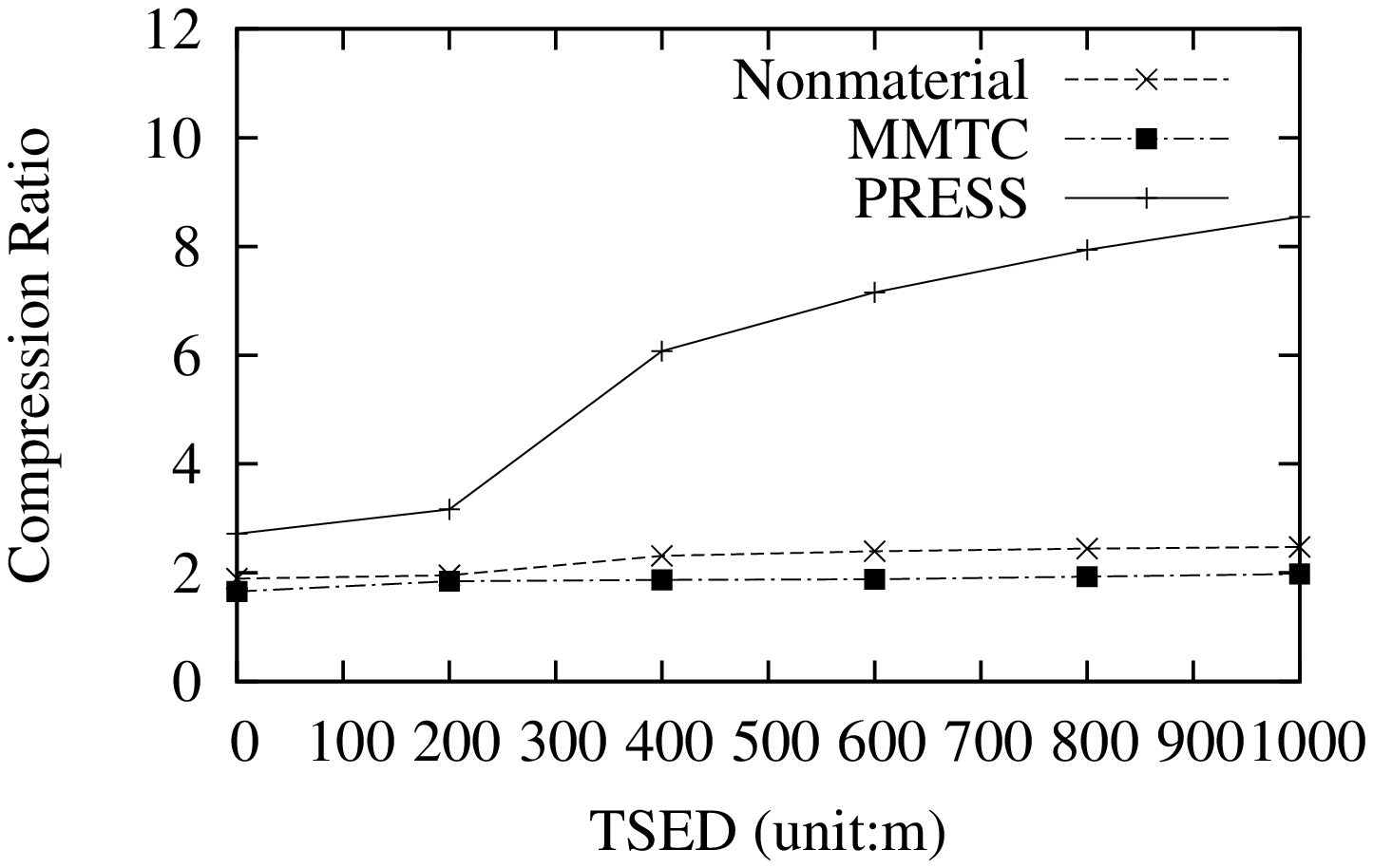,angle=270,width=1\textwidth}
        \vspace{-0.2in}
        \caption{\footnotesize Compression ratio performance}
        \label{fig:cr-comparison}
    \end{minipage}
    \begin{minipage}[t]{.4\textwidth}
        \centering
        \psfig{figure=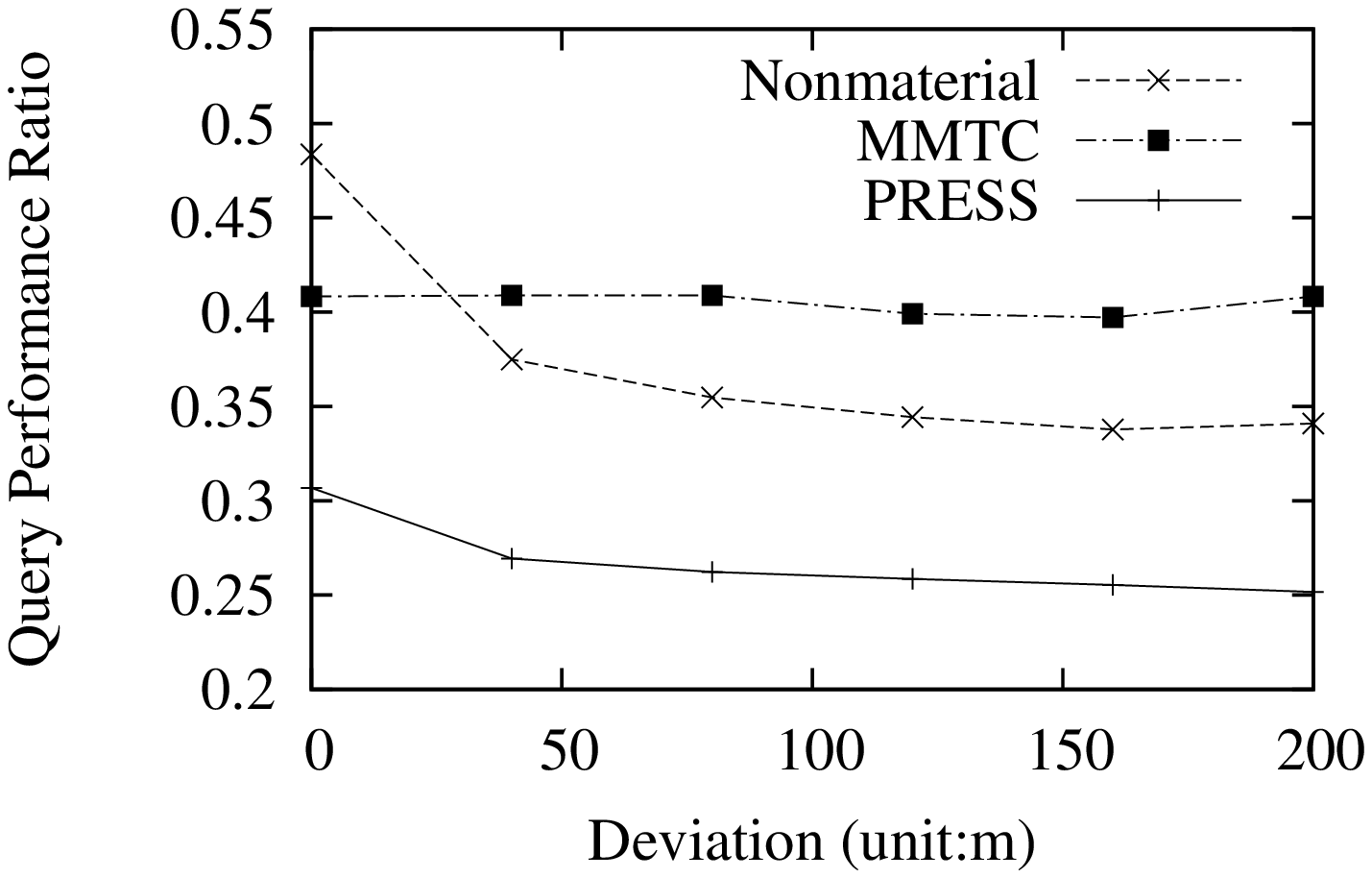,angle=270,width=1\textwidth}
        \vspace{-0.2in}
        \caption{\footnotesize Performance of $where_{at}$ query}
        \label{fig:whereat}
    \end{minipage}
    \vspace{-0.15in}
\end{figure}

Finally, we present the overall compression ratio of all three algorithms (i.e., our PRESS framework and two competitors) in Fig.~\ref{fig:cr-comparison}. Notice that existing works are not bounded by NSTD and TSND but only by TSED and hence the results are presented based on various TSED values. It is very obvious that our approach is able to achieve a much higher compression ratio than existing ones. The higher the TSED is, the more significant the advantage of PRESS is. For example, when TSED = 0, our approach can improve MMTC's compression ratio by $64\%$ and it can improve Nonmaterial's compression ratio by $43\%$; when TSED = 600m, our approach can improve MMTC's compression ratio by $280\%$ and Nonmaterial's compression ratio by $199\%$. We also use the standard ZIP and RAR algorithm to compress the trajectory. The compression ratio of ZIP is 2.09, and that of RAR is 3.78. Although both algorithms are lossless, the compressed trajectories lose their utility totally. That is to say, the compressed trajectories have to be fully decompressed before use. Moreover, the compression ratio of PRESS is better than that of ZIP even when TSND = NSTD = 0, and it beats RAR when TSED $>$ 230m. As a summary, our approach is very effective in terms of compressing trajectories, which is the main objective of the compression framework proposed in this work. The experimental results well demonstrate the power of our framework and hence justify the design.

\subsection{Compression Efficiency}

\begin{figure}[htb]%
\vspace{-0.1in}
\begin{center}
\subfigure[Compression]{
   \label{fig:com-time}
   \psfig{file=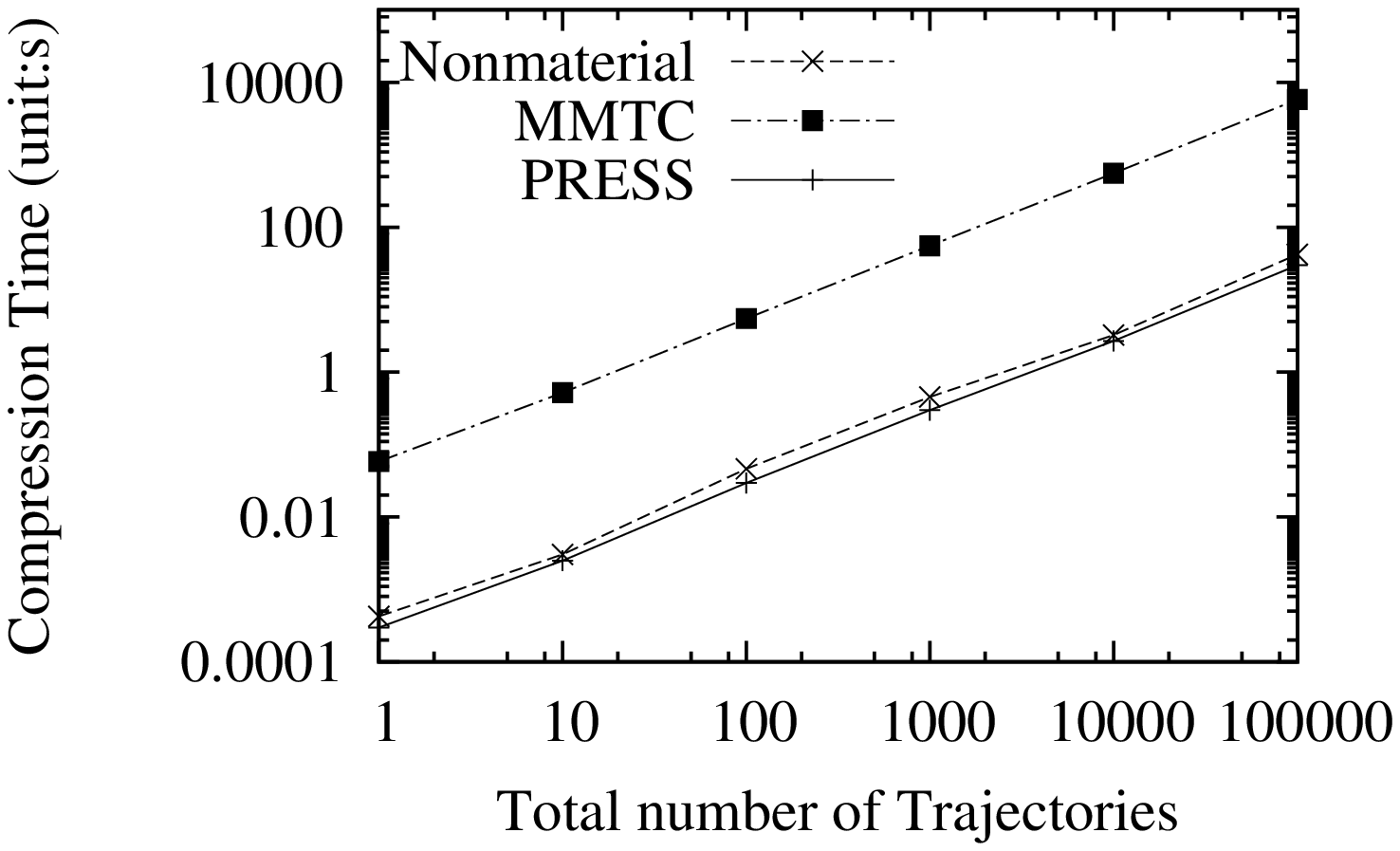,angle=270,width=0.4\textwidth}
   \vspace{-0.1in}
  }
\hspace{-0.1in}
\subfigure[De-compression]{
   \label{fig:dep-time}
   \psfig{file=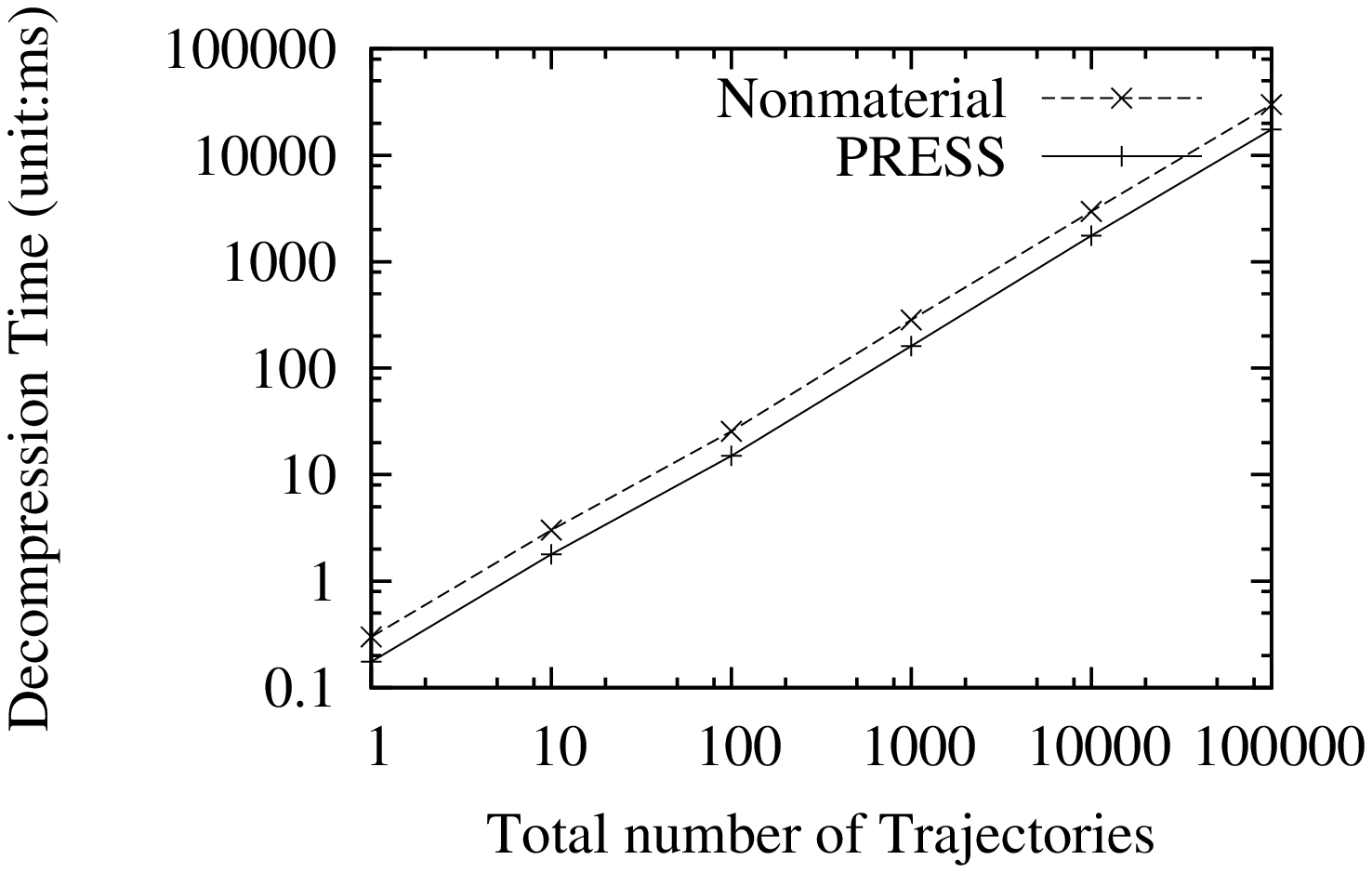,angle=270,width=0.4\textwidth}
   \vspace{-0.1in}
  }
\vspace{-0.15in}
\caption{Time efficiency}
\label{fig:time}
\end{center}
\vspace{-0.2in}
\end{figure}

Although the main objective of PRESS framework is to save space, the time complexity is also important. Ideally, we prefer a compression algorithm that can effectively cut down the storage cost and meanwhile is time efficient. Consequently, we report the time taken by each algorithm when compressing trajectories, and we also report the time taken by each algorithm for decompressing trajectories, as shown in Fig.~\ref{fig:time}. It is noticed that MMTC does not support decompression as the compressed trajectories cannot be recovered.

In terms of trajectory compression, MMTC is the most time-consuming. The average compression time cost of MMTC is 196.5 times of PRESS. Moreover, our approach is faster than Nonmaterial, and it on average only requires $72\%$ of Nonmaterial's time. Furthermore, PRESS also outperforms both ZIP and RAR in terms of time efficiency. It only takes around $24.6\%$ of ZIP's time and $13.0\%$ of RAR's time.
%
%
For decompression, our approach consumes $58.7\%$ time of Nonmaterial, $56.1\%$ time of ZIP, and $74.5\%$ time of RAR on average. The reason behind is that Nonmaterial contains arithmetic of real numbers to calculate the timestamps during decompression, while our approach only needs to visit the Huffman tree, Trie and shortest path table for recovery of the trajectories. Consequently, we can conclude that our approach is most time efficient, for both compression and decompression operations.

Some may argue that our approach relies on auxiliary structures, namely the shortest path table, the AC automation and the Huffman tree. Here, we want to point out that these structures are static and they are only constructed once. For the real dataset we use, they take space of 452MB, 101MB and 121MB respectively, which is absolutely acceptable for modern computation platforms. Compared with the storage saving they achieve for compressing trajectories, the cost of constructing these structures and maintaining these structures is well justified.

\subsection{Supporting LBS applications}

As mentioned before, our framework not only can effectively compress the trajectories, but also can support some popular spatial-temporal queries commonly used by many LBSs, even when the trajectories are in the compressed form. In the following, we report their time performance for supporting different queries.
For a given query $q$, an original trajectory dataset $TD$ and a compressed trajectory dataset $TD'$, let $t(q, TD)$ and $t(q, TD')$ represent the time duration taken by processing the query $q$ over the dataset $TD$ and $TD'$ respectively. We report the time performance ratio (i.e., $\frac{t(q, TD')}{t(q, TD)}$) instead of the time duration for each query.

First, we study the $where_{at}$ query, with its performance shown in Fig.~\ref{fig:whereat} under various distance deviations. In general, the time performance depends on the distance deviation. The higher the deviation, the higher the compression ratio and hence the faster the query processing. PRESS is most efficient, and it on average only takes $26\%$ of the time spent in original uncompressed trajectory dataset. Compared with its two competitors, PRESS saves around $34\%$ of MMTC's time and roughly $28\%$ of Nonmaterial's time.

\begin{figure}[htb]
\centering
    \begin{minipage}[t]{.4\textwidth}
        \centering
        \psfig{figure=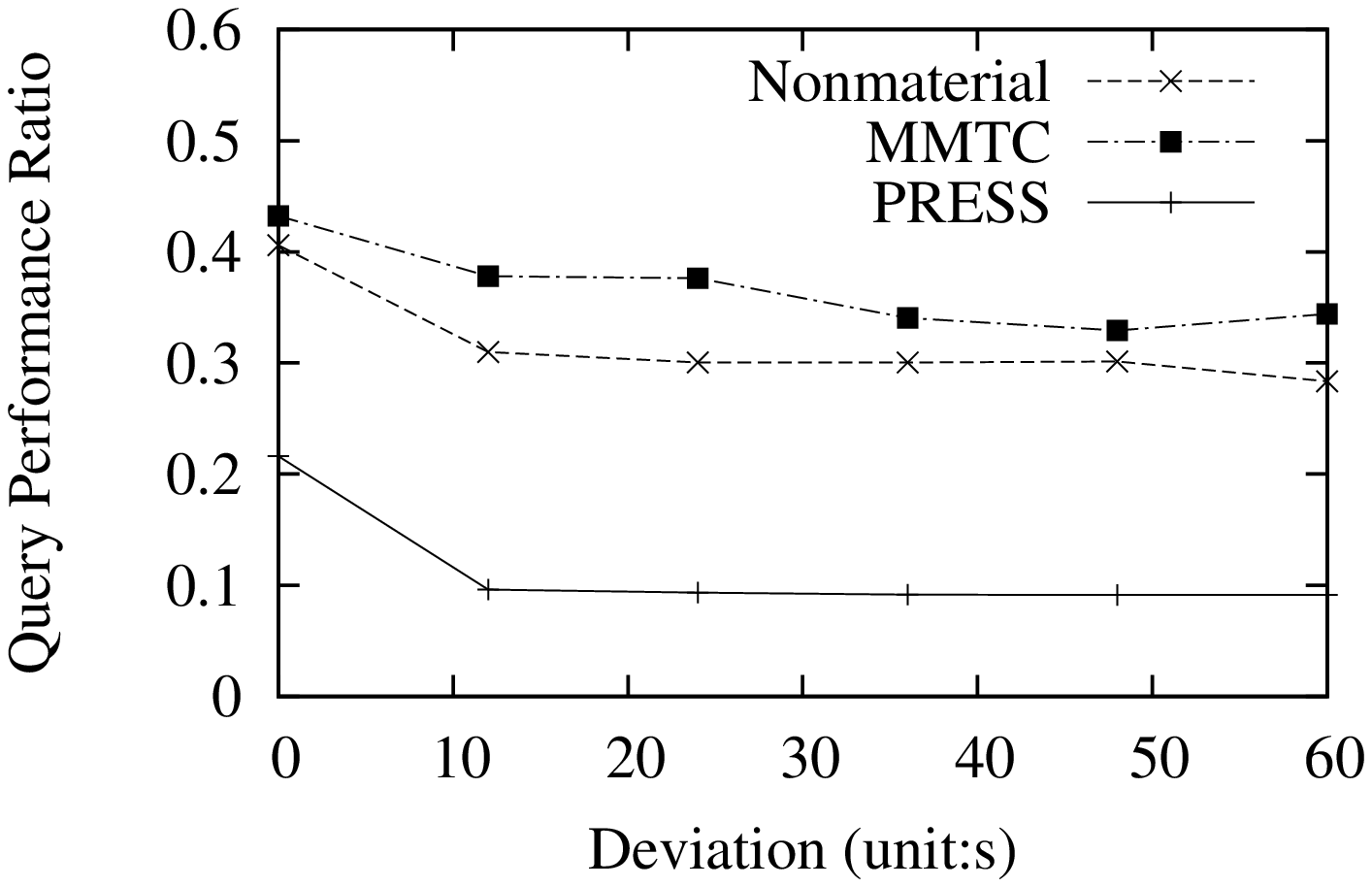,angle=270,width=1\textwidth}
        \vspace{-0.2in}
        \caption{\footnotesize Performance of $when_{at}$ query}
        \label{fig:whenat}
    \end{minipage}
    \begin{minipage}[t]{.4\textwidth}
        \centering
        \psfig{figure=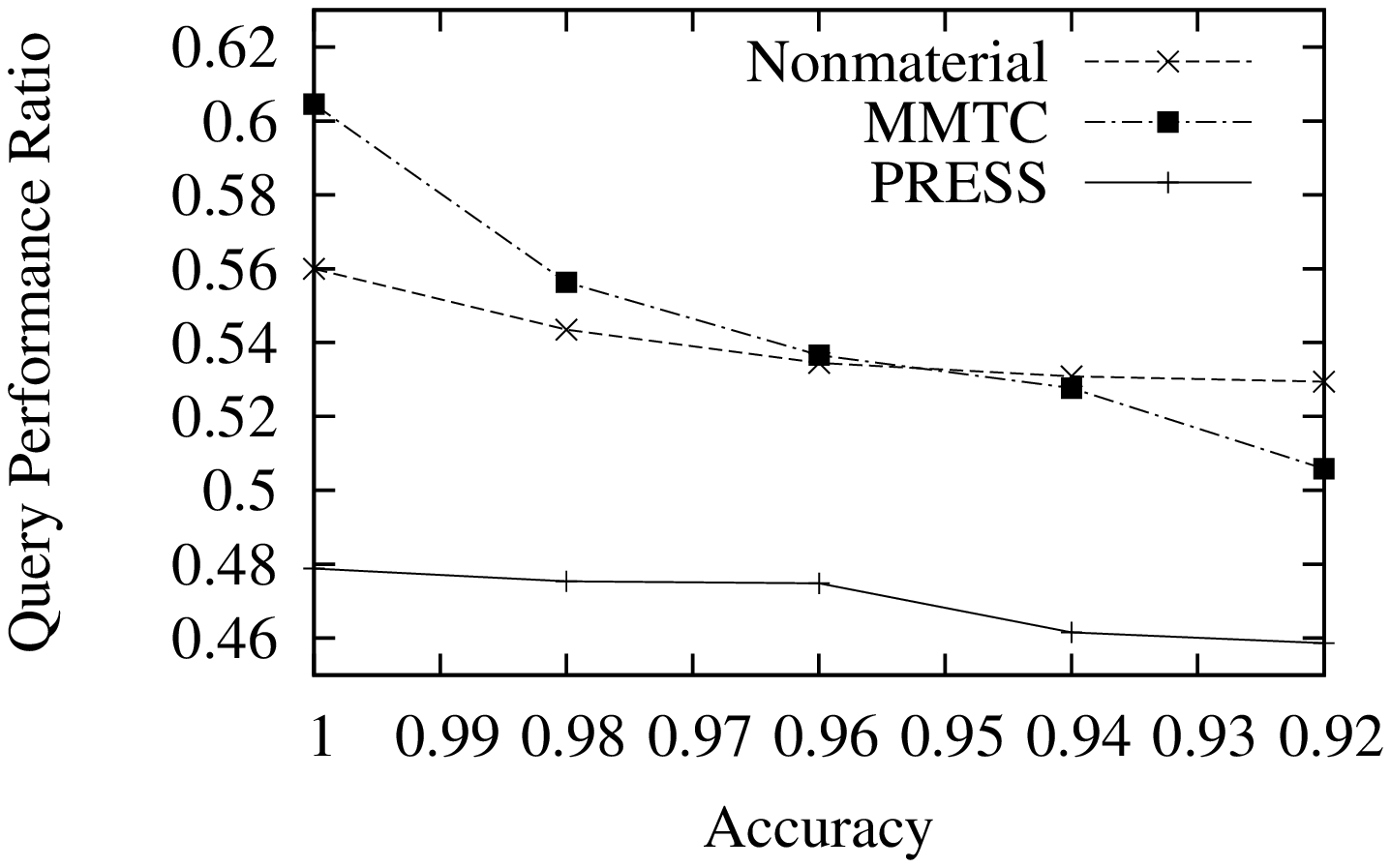,angle=270,width=1\textwidth}
        \vspace{-0.2in}
        \caption{\footnotesize Performance of $range$ query}
        \label{fig:rangequery}
    \end{minipage}
    \vspace{-0.15in}
\end{figure}

Second, we study the $when_{at}$ query, with its performance shown in Fig.~\ref{fig:whenat} under various time deviations. Again, our approach significantly outperforms the other two approaches. On average, it only incurs $30\%$ of MMTC's time and incurs around $35\%$ of Nonmaterial's time.

Then, we study the range query and report its performance under different accuracy in Fig.~\ref{fig:rangequery}. Unlike previous two queries, the range query returns a boolean value and it is not error bounded under any of the compression scheme studied in this evaluation. For example, when there are a large number of trajectories located near the boundary of the queried spatial range at the queried temporal period, the lossy temporal compression of our approach and Nonmaterial may result in wrong query result. On the other hand, MMTC is even worse, as it compresses the trajectories in such a way that it loses both the spatial information and the temporal information to certain degree. Consequently, we randomly generate $2,325,000$ range queries, and then cluster queries based on the accuracy. Again, PRESS works the best, and it can save around $14\%$ of the time, compared with both MMTC and Nonmaterial.

\noindent
\textbf{Discussion.} Although PRESS demonstrates excellent performance in all the above mentioned common spatial queries, it has to maintain certain auxiliary structures. If we only consider above three queries, our new framework needs to maintain the distance information of all-pair shortest path which takes around 904MB space, the distance information between Trie nodes which takes roughly 201MB space, the MBRs for all-pair shortest path which take 904MB space, and Trie MBR information which takes another 805MB space. However, given trajectories in a real road network which is relatively stable, the overhead of these auxiliary structures is acceptable, especially when these structures can significantly improve the query performance and they can be used for a long period. We want to highlight that for fair competition, we also maintain several auxiliary structures for MMTC and Nonmaterial. Note that the original papers of MMTC and Nonmaterial do not present any approaches to support spatial-temporal queries in the compressed form, and we have extended original work by adding extra structures in order to support the queries we studied in the paper. The original trajectory, however, does not need any auxiliary structure.

For the $where_{at}$ query, MMTC does not need any auxiliary structure. Nonmaterial needs a structure maintaining the distance of any consecutive part of a street. For $when_{at}$ query, MMTC needs a structure maintaining the MBRs for all the edges in the road network, and Nonmaterial needs a structure maintaining the MBRs of any consecutive part of a street. As a summary, these two approaches need less auxiliary space than PRESS, but this does not weaken the strength of our approach because of the superior compression power and its capability to accelerate many queries.

Note that all the methods we introduced in the paper and above do not utilize indexing structures. Since we have maintained the MBRs of edges, shortest paths and Trie nodes, we think PRESS is compatible to most, if not all, indexing structures such as R-tree and Quad-tree, and thus it can support more advanced queries using the existing works based on such indexing structures. We leave this as a future research direction.

\section{Related Work}
\label{sec:related}

In this section, we review related work to trajectory compression. They are roughly clustered into two groups, one to compress a trajectory using line simplification methods and the other based on map-matching algorithms.

\subsection{Trajectory Simplification}

Line simplification is to approximate a polyline with a subset of the vertices from the original one. These algorithms are used to compress Euclidean space trajectories. Existing line simplification algorithms can be categorized as either batch model based or online compression. According to an experimental study~\cite{GIS10_402} that compares the major compression approaches, no approach outperforms others under all scenarios as they have their pros and cons.

\subsubsection{Batch Model Based}

Those line simplification methods falling this category discard some locations with negligible error from an original trajectory which is already wholly obtained before the process~\cite{zheng2011computing}. The uniform sampling algorithm is an efficient (but not error-bounded) simplification method that keeps every $i$th points and discards others. The Douglas-Peucker (DP) algorithm~\cite{GIG73_112,hershberger1992speeding} approximates a trajectory by a line segment and recursively selects the point contributing the biggest error as a split point, until the trajectory satisfies the error requirement. Quite a few variants have been proposed to improve the original DP algorithm~\cite{AC86_103}, e.g., replacing the perpendicular distance with the time synchronized Euclidean distance or a time-distance ratio metric that considers both spatial and temporal information~\cite{EDBT04_765}, and a bottom-up algorithm that starts from the consecutive sample points and approximates step by step by merging the consecutive segments into one line segment which introduces the least error~\cite{KDD98_239}. The Bellman's algorithm~\cite{ACMC61} uses dynamic programming to minimize the ``area" between the original trajectory and the compressed one. However, the time complexities of the original DP method and Bellman's algorithm are $O(|T|^{2})$ and $O(|T|^{3})$ respectively. Some improved implementations of DP-variants can achieve a complexity of $O(|T|log|T|)$, and that of Bellman's algorithm can be $O(|T|^{2})$.

As compared with this category of algorithms, PRESS also works on batch model. In addition, the complexity of PRESS is $O(|T|)$, much more efficient than DP-variants and Bellman's algorithms. Recently, a Spatial Quality Simplification Heuristic Extended approach~\cite{GEO13_1} has been proposed with the time complexity of $O(|T|)$, achieving a relatively low error in Euclidean spaces. Different from this approach, PRESS works in road network spaces, in which we face more constrains. Furthermore, PRESS is spatial lossless and temporal error-bounded.

\subsubsection{Online compression}

Algorithms of this category compress the trajectories that are being generated, i.e., we need to make a decision whether the recently received point should be reserved or not online. The reservoir sampling algorithm~\cite{TOMS85_37} uses the replace strategy to keep no more than a maximal number of sample points to assure that each point shares the same probability to be kept at last. The sliding window approach~\cite{EDBT04_765,ICDM01_289}, on the other hand, tries to simplify the points within a sliding window with a line. The sliding window keeps growing until the simplification exceeds the error limitation. The sliding window approach also inspires other variations, such as the opening window approach~\cite{EDBT04_765} and Dead Reckoning~\cite{DEWMA06_19}. Other approaches argue that a point should be included in the compression result as long as it reveals significant change of the movement. Given speed and direction error bounds, the STTrace algorithm~\cite{SSDM06_275} uses the concept of safe area to generate a simplified trajectory. Alternatively, work presented in~\cite{EDBT04_765} explores the speed information and uses the speed difference of two sub-trajectories as an error metric to determine whether to reserve a sample point.

Although PRESS focuses on batch model based trajectory compression, the compression procedure scans the spatial path and temporal sequence from head to tail without tracing back. This means PRESS can be adapted to online compression. In BTC, we propose a novel angular-range approach with a time complexity of $O(|T|)$ based on a variant of the opening window method whose original complexity is $O(|T|^{2})$. Additionally, the trajectory compressed by PRESS can be directly used to answer spatial-temporal queries without being fully decompressed.

\subsection{Map-Matching Based Compression}

This fold of compression methods first projects a trajectory onto a road network using a map-matching algorithm~\cite{VLDB05_853,GIS09_352,GIS09_336,GIS12_605}, and then reduces the storage of the trajectory that has been represented as a sequence of road segments. Here, we review two representatives, Nonmaterial~\cite{ICDT05_173} and MMTC~\cite{JSS13}. Nonmaterial uses the street information to represent the spatial information. It calculates the timestamps of the intersections based on the timestamps of the original sampled points and the spatial locations of their snapped projections on the roads, under the assumption of uniform speed movement. MMTC uses sub-trajectories through fewer intersections to replace parts of the original trajectory. Some specific evaluation functions are introduced during the compression to guarantee the similarity between the compressed trajectory and the original one. The compressed trajectory consists of fewer intersections, thus the storage cost is reduced.

To the best of our knowledge, Nonmaterial and MMTC are the only techniques focusing on trajectory compression in road networks. Different from these two approaches, we compress the spatial component and temporal component of a trajectory separately. In the meantime, the spatial compression reduces the storage of a trajectory tremendously by using the sequential patterns mined from the trajectory corpus to encode the trajectory. As a result, our approach achieves a much higher compression ratio with a linear time complexity. According to the extensive experiments, PRESS outperforms both Nonmaterial and MMTC in terms of both effectiveness and efficiency.

\section{Conclusion}
\label{sec:conclusion}

In this paper, we propose a new framework, namely PRESS, for trajectory compression. To be more specific, PRESS represents a given trajectory using a spatial path and a temporal sequence and then employs different algorithms to compress the spatial path and temporal sequence, respectively. We have conducted extensive experiments to evaluate the performance of PRESS on a real data set, i.e., the Singapore taxi trajectories collected within one month. The simulation results demonstrate the superior compression power of PRESS. In the near future, we plan to formalize the selection of the $\theta$ value for FST compression, and we will collect more types of real trajectories (e.g., pedestrian trajectory) to evaluate how PRESS works in different application scenarios.

\section{Acknowledgments}
\label{sec:acknowledge}

This work is partially supported by the National Natural Science Foundation of China (NSFC) under grant No. 61073001 and Shanghai Leading Academic Discipline Project, Project Number: B114.

\bibliographystyle{plain}

\bibliography{paper-ref-short-new}

\end{document}